\documentclass[10pt, a4paper]{article}
\pdfoutput=1
\usepackage{jcappub}
\usepackage{amsmath,amssymb,graphicx,xspace,subfigure,tikz}
\usepackage{bm}
\usepackage{mathrsfs}
\usepackage{ulem}
\usepackage{hyperref} 
\usepackage{multirow}
\hypersetup{
    colorlinks=true,       % false: boxed links; true: colored links
    linkcolor=red,          % color of internal links
    citecolor=blue,        % color of links to bibliography
    filecolor=magenta,      % color of file links
    urlcolor=blue           % color of external links
}
\usepackage[all]{hypcap} 

\newcommand{\ba}[1]{\begin{align} #1 \end{align}}
\newcommand{\bes}[1]{\begin{equation}\begin{split} #1 \end{split}\end{equation}}
\newcommand{\bsa}[2]{\begin{subequations}\label{#1}\begin{align} #2 \end{align}\end{subequations}}
\newcommand{\com}{\;,}
\newcommand{\per}{\;.}
\newcommand{\cc}{\mathrm{c.c.}}

\newcommand{\nn}{\nonumber \\}

\newcommand{\fref}[1]{Figure~\ref{#1}}

\newcommand{\tref}[1]{table~\ref{#1}}
\newcommand{\Tref}[1]{Table~\ref{#1}}
\newcommand{\sref}[1]{section~\ref{#1}}

\newcommand{\aref}[1]{appendix~\ref{#1}}

\newcommand{\eref}[1]{Eq.~(\ref{#1})}
\newcommand{\erefs}[2]{Eqs.~(\ref{#1})~and~(\ref{#2})}

\newcommand{\rref}[1]{Ref.~\cite{#1}}
\newcommand{\rrefs}[1]{Refs.~\cite{#1}}
\newcommand{\epsF}{\epsilon_F}
\newcommand{\epsP}{\epsilon_+}
\newcommand{\epsM}{\epsilon_-}
\newcommand{\qe}{qe}
\newcommand{\Lscr}{\mathscr{L}}
\newcommand{\Fcal}{\mathcal{F}}
\newcommand{\Mcal}{\mathcal{M}}
\newcommand{\Ical}{\mathcal{I}}

\newcommand{\dd}{\mathrm{d}}
\newcommand{\ii}{\mathrm{i}}
\newcommand{\ee}{\mathrm{e}}
\newcommand{\Xbb}{\mathbb{X}}
\newcommand{\Ubb}{\mathbb{U}}
\newcommand{\Abb}{\mathbb{A}}
\newcommand{\Pbb}{\mathbb{P}}
\newcommand{\Rbb}{\mathbb{R}}

\newcommand{\IN}{\text{\sc in}}
\newcommand{\OUT}{\text{\sc out}}
\newcommand{\FREE}{\text{\sc free}}
\newcommand{\UV}{\text{\sc uv}}
\newcommand{\IR}{\text{\sc ir}}

\newcommand{\ket}[1]{\bigl|#1\bigr>}
\newcommand{\amp}[2]{\bigl< #1 \bigl| #2\bigr>}

\newcommand{\expval}[3]{\bigl< #1 \bigr| #2 \bigl| #3 \bigr>}

\newcommand{\xvec}{{\bm x}}

\newcommand{\pvec}{{\bm p}}
\newcommand{\phat}{\hat{\pvec}}
\newcommand{\qvec}{{\bm q}}

\newcommand{\Fvec}{{\bm F}}

\newcommand{\avec}{{\bm a}}

\newcommand{\kvec}{{\bm k}}

\newcommand{\vvec}{{\bm v}}
\newcommand{\vhat}{\hat{\vvec}}
\newcommand{\Xvec}{{\bm X}}

\hypersetup{
  pdfauthor={Andrew Long,Jessica Turner},
  pdftitle={Thermal pressure on ultrarelativistic bubbles from a semiclassical formalism},
  pdfkeywords={Parton Shower,Electroweak Phase Transition}
}

\begin{document}

\title{Thermal pressure on ultrarelativistic bubbles from a semiclassical formalism}

\author[a]{Andrew~J.~Long}
\affiliation[a]{Rice University, Houston, TX, 77005, U.S.A.}

\author[b]{and Jessica~Turner}
\affiliation[b]{Institute for Particle Physics Phenomenology, Durham University, Durham, U.K.} 

\preprintnumber{IPPP/24/50}  % Add your preprint number here

\abstract{
We study a planar bubble wall that is traveling at an ultrarelativistic speed through a thermal plasma.  This situation may arise during a first-order electroweak phase transition in the early universe.  As particles cross the wall, it is assumed that their mass grows from $m_a$ to $m_b$, and they are decelerated causing them to emit massless radiation ($m_c=0$).  We are interested in the momentum transfer to the wall, the thermal pressure felt by the wall, and the resultant terminal velocity of the wall.  We employ the semiclassical current radiation (SCR) formalism to perform these calculations.  An incident-charged particle is treated as a point-like classical electromagnetic current, and the spectrum of quantum electromagnetic radiation (photons) is derived by calculating appropriate matrix elements.  To understand how the spectrum depends on the thickness of the wall, we explore simplified models for the current corresponding to an abrupt and a gradual deceleration.  For the model of abrupt deceleration, we find that the SCR formalism can reproduce the $P_\mathrm{therm} \propto \gamma_w^0$ scaling found in earlier work by assuming that the emission is soft, but if the emission is not soft the SCR formalism can be used to obtain $P_\mathrm{therm} \propto \gamma_w^2$ instead.  For the model of gradual deceleration, we find that the wall thickness $L_w$ enters to cutoff the otherwise log-flat radiation spectrum above a momentum of $\sim \gamma_w^2 / L_w$, and we discuss the connections with classical electromagnetic bremsstrahlung. 
}


\maketitle

\setlength{\parindent}{20pt}
\setlength{\parskip}{2.5ex}

%==================================
% Introduction
%==================================
\section{Introduction}
\label{sec:introduction}

%=========
In the early universe, the primordial plasma cooled as a consequence of the cosmological expansion~\cite{Kolb:1990}.  
When the plasma cooled to a temperature commensurate with the electroweak energy scale, the Standard Model Higgs field developed a nonzero vacuum expectation value during an event known as the cosmological electroweak phase transition~\cite{Dine:1992wr}.  
Whereas the Standard Model predicts that this transition is a continuous crossover~\cite{Kajantie:1996mn,DOnofrio:2015mpa}, simple extensions of the Standard Model predict that it is first order instead~\cite{Espinosa:1993bs}, corresponding to the nucleation and percolation of Higgs-phase bubbles.  
The speed of these Higgs-phase bubble walls is of particular interest since their dynamics may be the origin of various observable cosmological relics including the matter-antimatter asymmetry of the universe~\cite{Cohen:1990it}, primordial magnetism~\cite{Vachaspati:1991nm}, and gravitational wave radiation~\cite{Kamionkowski:1993fg}. 
More generally, first-order phase transitions in a dark sector can also provide interesting cosmological signatures including dark matter~\cite{Dimopoulos:1990ai,Bai:2018dxf,Heurtier:2019beu,Baker:2019ndr,Chway:2019kft,Gehrman:2023qjn,Giudice:2024tcp}, primordial black holes \cite{Gross:2021qgx,Baker:2021nyl,Kawana:2021tde,Liu:2021svg,Baker:2021sno,Liu:2022lvz,Gouttenoire:2023naa,Cai:2024nln}, gravitational waves~\cite{Azatov:2019png}, particle production \cite{Shakya:2023kjf,Mansour:2023fwj}, and leptogenesis \cite{Pascoli:2016gkf,Long:2017rdo,Shuve:2017jgj,Chun:2023ezg,Cataldi:2024pgt}.

%=========
In general, the dynamics of a domain wall is governed by the forces acting upon it~\cite{Espinosa:2010hh}. 
A differential vacuum pressure across the phase boundary compels the bubbles to grow and accelerates the bubble wall outward.  
At the same time, a differential thermal pressure from interactions with the plasma inhibits the bubble's growth and tends to decelerate the bubble wall.  
Whereas a calculation of the vacuum pressure is fairly straightforward, a calculation of the thermal pressure is more subtle.  

%=========
The calculation of thermal pressure can be broadly divided into two regimes based on the magnitude of the bubble wall velocity, $\vvec_w$. 
When $|\vvec_w|$ is sufficiently small, a fluid description of the plasma can be applied. 
In this regime, the backreaction of particles in the plasma interacting with the bubble wall can be estimated by solving the Boltzmann equation for all particle species along with the equation of motion for the scalar field~\cite{Moore:1995ua, Moore:1995si,Espinosa:2010hh, Dorsch:2023tss, Cline:2020jre, Laurent:2020gpg, Dorsch:2021ubz, Dorsch:2021nje, Cline:2021iff, Cline:2021dkf, Lewicki:2021pgr, Laurent:2022jrs, Ellis:2022lft,DeCurtis:2022hlx,DeCurtis:2023hil}. 
This approach is challenging to employ, because it requires determining deviations from equilibrium distributions of different plasma species through the solution of the corresponding Boltzmann equations. 
Nonetheless, it has been shown that purely equilibrium hydrodynamic backreaction can hinder accelerating expansion, leading to a simple estimate that serves as an upper limit on bubble-wall velocity~\cite{BarrosoMancha:2020fay,Balaji:2020yrx,Ai:2021kak,Wang:2022txy,Ai:2023see,Wang:2023kux}. 
These findings assume that plasma profiles are always fully developed into steady-state solutions at a given wall velocity, a premise verified via numerical simulation assuming local thermal equilibrium~\cite{Krajewski:2024gma}.

%=========
In the regime where the bubble wall is highly relativistic, the calculation simplifies somewhat, because particles interact with the wall on a time scale that is short compared to the scattering time scale in the plasma.  
Over the last several years, a considerable body of work has developed to study the dynamics of ultrarelativistic bubbles and their potential signatures~\cite{Bodeker:2009qy,Bodeker:2017cim,Azatov:2020nbe,Hoche:2020ysm,Gouttenoire:2021kjv,Azatov:2023xem,Azatov:2024auq}.
For ultrarelativistic bubble walls, the thermal pressure (in the rest frame of the wall) has the general schematic form: $P_\mathrm{therm} \approx \Fcal_a \, \langle \Delta p_z \rangle$.  
Here $\Fcal_a \approx \gamma_w T^3$ is the typical flux of incident particles onto the wall assuming that these particles are drawn from a thermal bath at temperature $T$ and boosted with Lorentz factor $\gamma_w \gg 1$ into the rest frame of the wall.  
So the calculation of thermal pressure can be recast as a calculation of $\langle \Delta p_z \rangle$, the average longitudinal momentum transfer from the incident particles to the wall.  
Here we briefly summarize several key studies over the last several years.  

%=========
The authors of \rref{Bodeker:2009qy} (hereafter BM09) calculated the momentum transfer that results when a particle with energy $E_a$ is incident on a bubble wall at which its mass increases from $m_a$ to $m_b$.  
Since energy and transverse momentum are conserved, the restrictive kinematics of this ``1-to-1 transition'' implies that the longitudinal momentum decreases by $\Delta p_z \approx (m_b^2 - m_a^2) / 2E_a$ where $E_a$ is the energy of the incoming particle and is assumed to be large compared to the masses.  
If the particle is drawn from a plasma at temperature $T$, and if the wall travels with speed $v_w = |\vvec_w|$ relative to the plasma, then $E_a \sim \gamma_w T$ where $\gamma_w = (1 - v_w^2)^{-1/2}$. 
The Lorentz factor is large ($\gamma_w \gg 1$) for an ultrarelativistic wall, and it follows that $\Delta p_z \propto \gamma_w^{-1}$ and $P_\mathrm{therm} \propto \gamma_w^0$.  
Since the vacuum pressure is also insensitive to the bubble's speed, $P_\mathrm{vac} \propto \gamma_w^0$, BM09 concluded that Higgs-phase bubble walls might accelerate without bound and ``runaway'' if $P_\mathrm{vac} > P_\mathrm{therm}$.  

%=========
Several years later, the authors of BM09 revisited the problem of electroweak bubble wall velocity in \rref{Bodeker:2017cim} (hereafter BM17).  
Here, they extended their earlier analysis to account for ``1-to-2 transitions'' in which an incident particle emits soft radiation upon crossing through the wall.  
The presence of a third particle in the system changes the kinematics, and the calculation of $\langle \Delta p_z \rangle$ entails an averaging over the quantum emission probability.  
To account for the mass profile of the particles at the wall, the authors employed the leading-order WKB approximation to calculate the matrix elements for a single emission.  
The thermal pressure was found to be largest for massive vector boson emission (if this channel is available), since the momentum transfer is controlled by the mass of the radiation, implying $\langle \Delta p_z \rangle \propto m_c \propto \gamma_w^0$ and $P_\mathrm{therm} \propto \gamma_w^1$. 
(BM17 also note that the $\gamma_w^1$ scaling is absent for massless vector boson emission; \textit{i.e.}, $P_\mathrm{therm} \propto \gamma_w^0$ if $m_c = 0$.)
As the wall accelerates, $\gamma_w$ increases, this causes $P_\mathrm{therm} \propto \gamma_w^1$ to increase until it equals $P_\mathrm{vac}$. At this point, the wall travels with a terminal velocity, which is typically ultrarelativistic ($\gamma_w \gg 1$).  
However, depending on the parameters, walls may collide before the terminal velocity is reached; see \rref{Athron:2022mmm} for a discussion of strongly supercooled first-order phase transitions.  

%=========
The authors of \cite{Hoche:2020ysm} (hereafter HKLTW20) set out to extend the analysis of BM17 by resumming multiple soft emissions. 
They assessed that this calculation would be intractable in the formalism used by BM17, so the authors employed a semiclassical formalism instead. 
The semiclassical formalism allows for the resummation of soft radiation, as a Sudakov form factor can be constructed from first principles, ensuring the cancellation of poles emerging from the real and virtual corrections \cite{Gribov:1983ivg,Bassetto:1983mvz}. 
This approach conserves probability and accurately describes the observed radiation emission patterns from a wide variety of experiments \cite{Lonnblad:1992tz}.
In the semiclassical approach, the radiator particle is treated as a classical source, and matrix elements for the emission of quantum radiation are calculated. 
The authors use this approach to describe the deceleration of incoming particles as they encounter the bubble wall, to calculate the corresponding radiation pattern, and to calculate the average longitudinal momentum transfer to the wall $\langle \Delta p_z \rangle$. 
Additionally, the authors treated the bubble wall as having zero thickness, making the deceleration of the classical source arbitrarily abrupt. 
Using this approach, they found that the spectrum of radiation was log-flat up to an ultraviolet (UV) cutoff set by the incident particle's energy $E_a$. 
Consequently, $\langle \Delta p_z \rangle \sim E_a \propto \gamma_w^1$ and $P_\mathrm{therm} \propto \gamma_w^2$. 
We review and discuss this calculation in \sref{sec:model2}. 
It is worth noting that the authors of HKLTW20 neglect the mass of the radiated particle in their analysis, \textit{i.e.} taking $m_c = 0$, but since $\langle \Delta p_z \rangle \propto \gamma_w T \gg m_c$ they argue that their results should carry over to massive emission. 

%=========
In this article, we clarify and extend the calculation in HKLTW20, having two goals in mind.  
First, we seek to clarify how the semiclassical current radiation (SCR) formalism used by HKLTW20 led them to conclude $P_\mathrm{therm} \propto \gamma_w^2$.  
We do so in a model with massless vector boson emission ($m_c = 0$).  
For comparison, the quantum particle splitting (QPS) formalism used by BM17 gives $P_\mathrm{therm} \propto \gamma_w^0$ for massless vector boson emission. 
We will show that the SCR formalism may be used to derive either $P_\mathrm{therm} \propto \gamma_w^2$ or $P_\mathrm{therm} \propto \gamma_w^0$, and that the essential ambiguity regards the choice of the radiating particle's momentum inside the bubble ($\pvec_b$).  
Second, we seek to extend the calculation in HKLTW20 to account for the finite thickness of the bubble wall.  
As is familiar from studies of deep inelastic scattering, the finite size of the target enters as a UV cutoff on the momentum distribution.  
Similarly, here one expects the finite wall thickness $L_w$ to cut-off the distribution~\cite{Bodeker:2017cim}.  
For typical models of a first-order electroweak phase transition the parameters are such $L_w^{-1} \ll \gamma_w T$.  
This suggests that there should be a large change in the thermal pressure when the finite wall thickness is taken into account~\cite{BarrosoMancha:2020fay,Azatov:2020ufh,Gouttenoire:2021kjv}.  

%=========
The remainder of this article is organized as follows.  
In \sref{sec:formalism} we review the semiclassical current radiation formalism, which provides a framework for calculating quantum electromagnetic radiation (photon emission spectrum) from classical currents.  
In sections~\ref{sec:vacuum}~and~\ref{sec:emission} we apply this formalism to derive expressions for the vacuum persistence probability and the single photon emission probability in terms of an arbitrary classical current density.  
In sections~\ref{sec:model1},~\ref{sec:model2},~and~\ref{sec:model3} we study three different ``models'' for the classical current.  
These correspond to a charged point-like particle traveling with constant velocity, a particle that decelerates abruptly, and a particle that decelerates gradually. 
For each model we calculate the spectrum of electromagnetic radiation and average longitudinal momentum transfer.  
Finally in \sref{sec:conclusion} we summarize our key results and discuss their implications for studies of Higgs-phase bubble wall velocities at a first-order cosmological electroweak phase transition.  
The article contains two appendices: \aref{app:cancellation} in which we remark on the cancellation of IR divergences and \aref{app:comparison} in which we review the QPS formalism developed by BM17.  

%==================================
% Semiclassical current radiation formalism
%==================================
\section{Semiclassical current radiation formalism}
\label{sec:formalism}

%=========
In the semiclassical current radiation (SCR) formalism, the incident and recoiling charged particles are modelled using a point-like classical electromagnetic current density $j^\mu(x)$.  As the particle decelerates, there is a probability that it emits an arbitrary number of photons of arbitrary momentum. This probability can be calculated by evaluating the corresponding matrix elements.  
In this section, we review the formalism, and in the following two sections, we use it to calculate probabilities.  

%-------------------------------------------
% Point-like classical particle
%-------------------------------------------
\subsection{Point-like classical particle}
\label{sub:particle}

\begin{figure}[t]
\centering \includegraphics[width=0.5\textwidth]{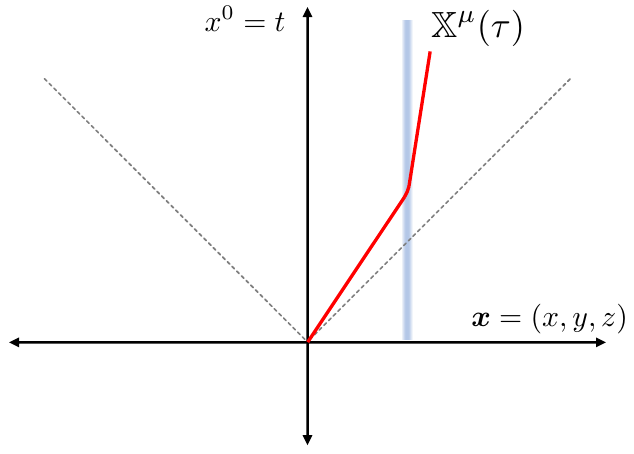}
\caption{An illustration of the worldline of the charged particle (red curve).  The particle initially travels with velocity $\vvec_a$, it encounters a domain wall (blue band), and it decelerates to velocity $\vvec_b$ with $|\vvec_b| < |\vvec_a|$.  }
\label{fig:worldline}
\end{figure}

%=========
Consider a point-like classical particle.  
Its location is given by the worldline $\Xbb^\mu(\tau)$, which is a map from the particle's private clock (proper time $\tau$) to a point in spacetime, represented by a position four-vector,   
see \fref{fig:worldline}.  
The corresponding velocity and acceleration four-vectors are calculated as 
\ba{\label{eq:Umu}
	\Ubb^\mu(\tau) = \frac{\dd \Xbb^\mu(\tau)}{\dd \tau} 
	\qquad \text{and} \qquad 
	\Abb^\mu(\tau) = \frac{\dd^2 \Xbb^\mu(\tau)}{\dd \tau^2} 
	\per
}
We require $\Ubb^0(\tau) > 0$ for all $\tau$, so the particle only moves forward in time.  
We take the metric for Minkowski spacetime to be $\eta_{\mu\nu} = \mathrm{diag}(1,-1,-1,-1)$, and we normalize the proper time $\tau$ such that
\ba{\label{eq:U_norm}
	\Ubb(\tau) \cdot \Ubb(\tau) = 1\,,
}
for all $\tau$.  
By taking the time derivative of the normalization condition, it follows that $\Ubb(\tau) \cdot \Abb(\tau) = 0$ for all $\tau$. 
In a frame of reference where the particle is moving with three-velocity $\vvec(\tau)$ we can write 
\bes{\label{eq:Umu_to_v}
	\Xbb^\mu(\tau) & = \bigl( T(\tau) \, , \, \Xvec(\tau) \bigr) \\ 
	\Ubb^\mu(\tau) & = \bigl( \gamma(\tau) \, , \, \gamma(\tau) \, \vvec(\tau) \bigr) \\ 
	\Abb^\mu(\tau) & = \bigl( \gamma^\prime(\tau) \, , \, \gamma^\prime(\tau) \, \vvec(\tau) + \gamma(\tau) \, \vvec^\prime(\tau) \bigr) 
	\com
}
where $\gamma(\tau) = 1 / \sqrt{1 - |\vvec(\tau)|^2}$ is the Lorentz factor and prime denotes derivative with respect to $\tau$.

%=========

%-------------------------------------------
% Varying particle mass
%-------------------------------------------
\subsection{Varying particle mass}
\label{sub:mass}

%=========
To use the SCR formalism to calculate electromagnetic radiation, it is only necessary to specify the particle's worldline $\Xbb^\mu(\tau)$.  
It is not necessary to specify what force is pushing or pulling on the particle, causing it to accelerate and radiate.  
However, it is useful to imagine that the particle is being accelerated when it enters a region of space (such as the bubble wall) where its mass is varying.   For this purpose, we can promote the particle's mass to be a four-scalar field $m(x)$.  
However, we are not interested in the dynamics of the mass field, but instead, we take this to be a fixed background, and we study the dynamics of particles that are propagating around in this region of space.  

%=========
The action for a point particle with variable mass $m(x)$ in Minkowski spacetime is written as
\ba{
	S[\Xbb(\tau)] = \int \! \dd \tau \, L\bigl(\Xbb(\tau), \, \dot{\Xbb}(\tau) \bigr) 
	\qquad \text{with} \qquad 
	L(x, u) = - \sqrt{u \cdot u} \, m(x)
	\com
}
which is a functional of the worldline $\Xbb^\mu(\tau)$.  
The mass field is only evaluated on the worldline.  
Varying the action with respect to each component of the worldline leads to the Euler-Lagrange equation 
\ba{
	\frac{\dd}{\dd \tau} \frac{\partial L}{\partial u^\mu} - \frac{\partial L}{\partial x^\mu} \biggr|_{x = \Xbb(\tau), \, u = \Ubb(\tau)} = 0 
	\per
}
After also using \eref{eq:U_norm}, one arrives at the equation of motion
\ba{\label{eq:mA}
    m(\Xbb) \, \Abb^\mu & = 
	\Bigl( \eta^{\mu\nu} - \Ubb^\mu \Ubb^\nu \Bigr) \frac{\partial m}{\partial x^\nu}  \biggr|_{x = \Xbb(\tau)} 
	\per
}
This is the analog of $m \avec = \Fvec$ .  
The quantity on the right-hand side of \eref{eq:mA} is interpreted as the four-force arising from the varying mass.  
By contracting this equation with $\Ubb^\mu$, we can show that the evolution is consistent with $\Ubb \cdot \Abb = 0$.  

%=========
To discuss the kinematics of the radiation, it is useful to define the four-momentum as 
\ba{
	\Pbb^\mu(\tau) = m\bigl(\Xbb(\tau)\bigr) \, \Ubb^\mu(\tau) 
	\per
}
The normalization of the four-velocity from \eref{eq:U_norm} implies 
\ba{
	\Pbb(\tau) \cdot \Pbb(\tau) = \bigl[ m\bigl(\Xbb(\tau)\bigr) \bigr]^2 
	\com
}
so that at any point along the worldline, the particle's four-momentum squares to its mass at that location.  
We can also evaluate the derivative 
\ba{
    \frac{\dd \Pbb^\mu}{\dd \tau} 
	= \Bigl( \Abb^\mu + \Ubb^\mu \Ubb^\nu \frac{\partial}{\partial x^\nu} \Bigr) m(x) \biggr|_{x = \Xbb(\tau)}
	\per
}
When combined with \eref{eq:mA}, the equation of motion can be written as 
\ba{
    \frac{\dd \Pbb^\mu}{\dd \tau} 
	= \eta^{\mu\nu} \frac{\partial m}{\partial x^\nu} \biggr|_{x = \Xbb(\tau)}
	\per
}
This is the analog of $\dd \pvec / \dd t = \Fvec$.  
Note that the force appearing here and the force in the analog of $\Fvec = m \avec$ from \eref{eq:mA} are not the same. 

%-------------------------------------------
% Electromagnetic current
%-------------------------------------------
\subsection{Electromagnetic current}
\label{sub:EM_current}

%=========
Suppose that the particle carries an electromagnetic charge $\qe$ where $q$ is dimensionless and $e = \sqrt{4 \pi \alpha} \approx 0.303$ is the electric charge of a proton in the Heaviside-Lorentz unit system.  
The corresponding electromagnetic current density may be calculated as 
\ba{\label{eq:jmu}
	j^\mu(x) = \qe \int_{-\infty}^{\infty} \! \dd\tau \, \Ubb^\mu(\tau) \, \delta^{(4)}\bigl( x - \Xbb(\tau) \bigr) 
 \com
}
where the integral runs over the particle's whole worldline, assuming it exists for all time.  
Some expressions later can be expressed more compactly in Fourier space.
The Fourier transform and inverse Fourier transform of the current density are defined by the relations 
\ba{\label{eq:Fourier_transform}
	j^\mu(x) = \int \! \! \frac{\dd^4 k}{(2\pi)^4} \, \tilde{j}^\mu(k) \, \ee^{- \ii k \cdot x} 
	\qquad \text{and} \qquad 
	\tilde{j}^\mu(k) = \int \! \dd^4 x \, j^\mu(x) \, \ee^{\ii k \cdot x} 
	\com
}
where $k^\mu = (k^0, \kvec)$ is called the wavevector.  
Using \eref{eq:jmu} gives 
\ba{\label{eq:jmuFT}
	\tilde{j}^\mu(k) = \qe \int_{-\infty}^{\infty} \! \dd\tau \, \Ubb^\mu(\tau) \, \ee^{\ii k \cdot \Xbb(\tau)} 
	\per
}
Since electromagnetic charge is conserved, the current must satisfy the continuity equation $\partial_\mu j^\mu(x) = 0$ for all $x^\mu$, or equivalently $k \cdot \tilde{j}(k) = 0$ for all $k^\mu$.  

%-------------------------------------------
% Electromagnetic interaction
%-------------------------------------------
\subsection{Electromagnetic interaction}
\label{sub:EM_interaction}

%=========
We now introduce the electromagnetic field $\hat{A}_\mu(x)$, which is a Hermitian quantum operator.  
We use hats to denote quantum operators.  
The Hamiltonian for this system takes the form 
\ba{
	\hat{H}(t) = \hat{H}_0 + \hat{H}_\mathrm{int}(t) 
	\com
}
where the free part of the Hamiltonian follows from the usual EM Lagrangian $\Lscr = - F_{\mu\nu} F^{\mu\nu}/4$, and the interaction part of the Hamiltonian is
\ba{\label{eq:Hint}
	\hat{H}_\mathrm{int}(t) = \int \! \dd^3 \xvec \ j^\mu(t,\xvec) \, \hat{A}_\mu(t,\xvec) 
	\per
}
We work in the interaction picture where operators evolve according to Hamilton's equation induced by $\hat{H}_0$ and states evolve according to Schr\"odinger's equation induced by $\hat{H}_\mathrm{int}(t)$.  

%-------------------------------------------
% Photon field operator
%-------------------------------------------
\subsection{Photon field operator}
\label{sub:photon_operator}

%=========
The photon field operator $\hat{A}_\mu(x)$ obeys Hamilton's equations of motion induced by $\hat{H}_0$.  
As such, it can be decomposed onto a basis of plane waves as~\cite{Peskin:1995ev}
\ba{\label{eq:A_from_a}
	\hat{A}_\mu(x) = \int \! \! \frac{\dd^3 \pvec}{(2\pi)^3} \frac{1}{\sqrt{2E_p}} \sum_{s = \pm 1} \biggl[ \hat{a}_{\pvec,s} \, \varepsilon_\mu(\pvec,s) \, \ee^{- \ii p \cdot x} + \hat{a}_{\pvec,s}^\dagger \, \varepsilon_\mu^\ast(\pvec,s) \, \ee^{\ii p \cdot x} \biggr] 
	\com
}
where $p^\mu \equiv (E_p, \pvec)$ and $E_p \equiv |\pvec|$.  
The ladder operators, $\hat{a}_{\pvec,s}$ and $\hat{a}_{\pvec,s}^\dagger$, obey the usual algebra of commutation relations.  
The form of the polarization vectors $\varepsilon_\mu(\pvec,s)$ depends on the choice of gauge.  
The $\FREE$ vacuum is defined as the state that is annihilated by all of the lowering operators 
\ba{
	\hat{a}_{\pvec,s} \ket{0_\FREE} = 0 
	\qquad \text{for all $\pvec$ and $s$} 
	\per
}
The $\FREE$ one-particle states are constructed as 
\ba{\label{eq:ps_FREE}
	\ket{(\pvec,s)_\FREE} = \sqrt{2E_p} \, \hat{a}_{\pvec,s}^\dagger \, \ket{0_\FREE} 
	\per
}
Note that $\ket{0_\FREE}$ has mass dimension $0$, and $\ket{(\pvec,s)_\FREE}$ has mass dimension $-1$. 

%=========
The photon Feynman propagator is defined by
\ba{\label{eq:DF_def}
	D_{F,\mu\nu}(x-y) = \expval{0_\FREE}{\mathrm{T}\Bigl\{ \hat{A}_{\mu}(x) \, \hat{A}_{\nu}(y) \Bigr\} }{0_\FREE} \,,
}
where $\mathrm{T}\{ \cdot \}$ denotes time ordering (later to the left).
Writing out the time-ordered product gives
\bes{
	D_{F,\mu\nu}(x-y) 
	& = \expval{0_\FREE}{\hat{A}_{\mu}(x) \, \hat{A}_{\nu}(y)}{0_\FREE} \, \Theta(x^0 - y^0) 
	\\ & \quad 
	+ \expval{0_\FREE}{\hat{A}_{\nu}(y) \, \hat{A}_{\mu}(x)}{0_\FREE} \, \Theta(y^0 - x^0) \,,
}
where $\Theta(z)$ is the unit step function.  
Using the plane wave decomposition of the field operator, the Feynman propagator can also be written as 
\bes{
	D_{F,\mu\nu}(x-y) 
%---
	& = \int \! \! \frac{\dd^3 \pvec}{(2\pi)^3} \, \frac{1}{2 E_p} \, \biggl[ \Theta(x^0 - y^0) \, \ee^{- \ii p \cdot (x-y)} \, C_{\mu\nu}(\pvec) 
	+ \Theta(y^0 - x^0) \, \ee^{\ii p \cdot (x-y)} \, C_{\nu\mu}(\pvec) \biggr] \,,
}
where we have defined 
\ba{\label{eq:C_def}
	C_{\mu\nu}(\pvec) = \sum_{s=\pm1} \varepsilon_\mu(\pvec,s) \, \varepsilon_\nu^\ast(\pvec,s) 
	\per
}
The tensor $C_{\mu\nu}(\pvec)$ depends on the choice of gauge.  
For instance, in the Feynman gauge $C_{\mu\nu}(\pvec) = - g_{\mu\nu}$, and the Feynman propagator takes the form 
\ba{\label{eq:DF_Feynman}
	D_{F,\mu\nu}(x-y) 
%---
	& = - g_{\mu\nu} \, \int \! \! \frac{\dd^3 \pvec}{(2\pi)^3} \, \frac{1}{2 E_p} \, \biggl[ \Theta(x^0 - y^0) \, \ee^{- \ii p \cdot (x-y)} + \Theta(y^0 - x^0) \, \ee^{\ii p \cdot (x-y)} \biggr] 
	\per
}
This is just a factor of $(- g_{\mu\nu})$ multiplied by the usual Klein-Gordon Feynman propagator.  
By introducing another integration variable, one can write the photon Feynman propagator in a manifestly Lorentz-covariant form
\ba{\label{eq:DF_Fourier}
	D_{F,\mu\nu}(x-y) 
%---
	& = \lim_{\epsF \to 0^+} \bigl( - g_{\mu\nu} \bigr) \int \! \! \frac{\dd^4 k}{(2\pi)^4} \, \frac{\ii}{k \cdot k + \ii \epsF} \, e^{- \ii k \cdot (x-y)} \,,
}
where $k^\mu = (k^0, \kvec)$.  
The integral is performed along the real $k^0$ axis, and the regulator $\epsilon_F$ is introduced to provide the correct prescription for integrating around the poles at $k^0 = \pm |\kvec|$.  

%=========
Here we pause to emphasize a point about gauge dependence.  
In this study, we consider the deceleration of a massive charged radiator and the emission of massless photon radiation.  
More generally, we expect our results to generalize to any QED- or QCD-like theory.
We focus on this model in part because it offers a technical simplification.  
Since the photon is massless everywhere in the system (\textit{i.e.}, on both sides of the bubble wall), we expect observables to remain invariant under gauge transformations.  
This gauge freedom allows us to select the Feynman gauge.  
By working with a massless photon, we avoid complications that would arise if the vector boson's mass were changing across the bubble wall.  
See \rref{Azatov:2023xem} for an analysis of thermal friction in models with spontaneously broken gauge symmetry, and a discussion of how the degrees of freedom rearrange at the bubble wall. 

%-------------------------------------------
% Thermal pressure
%-------------------------------------------
\subsection{Thermal pressure}
\label{sub:pressure}

%=========
The SCR formalism may be used to calculate the spectrum of electromagnetic radiation that arises when the charged particle is accelerated at the bubble wall.  
The associated momentum transfer from the wall to the particle and radiation corresponds to a force on the wall that tends to retard its motion.  
Since the incident charged particle is assumed to be drawn from a thermal bath, we talk about the thermal pressure on the wall.  
In the rest frame of the bubble wall, the thermal pressure may be calculated as~\cite{Bodeker:2009qy} 
\ba{\label{eq:Ptherm}
    P_\mathrm{therm} = \nu_a \int \! \! \frac{\dd^3 \pvec_a}{(2\pi)^3} \, f_a(\pvec_a) \, v_{a,z} \, \langle \Delta p_z \rangle
    \com
}
where $a$ labels the incident particle, $\nu_a$ counts the number of internal degrees of freedom (\textit{e.g.}, spin, color, particle/antiparticle), $f_a(\pvec_a)$ is the one-particle phase space distribution function (assumed equal for all of $a$'s internal degrees of freedom), $v_{a,z} = p_{a,z} / E_a $ is the component of velocity in the direction normal to the planar bubble wall, and $\langle \Delta p_z \rangle$ is the average longitudinal momentum transfer.  

%=========
Of particular interest is how the thermal pressure depends on the velocity of the wall $\vvec_w$ (\textit{i.e.}, the relative velocity of the wall rest frame and the plasma rest frame).  
If the factor of $\langle \Delta p_z \rangle$ were absent, the momentum integral would simply give the longitudinal flux $\Fcal_{a,z}$ of species-$a$ particles in the rest frame of the wall.  
For an ultrarelativistic wall traveling through the plasma with Lorentz factor $\gamma_w = 1 / \sqrt{1 - |\vvec_w|^2}$, this flux is parametrically $\Fcal_{a,z} \sim \gamma_w T^3$ since the direction normal to the wall is Lorentz contracted.  
To track the parametric scalings, it is useful to write 
\ba{\label{eq:Ptherm_scaling}
    P_\mathrm{therm} \sim \gamma_w T^3 \, \langle \Delta p_z \rangle 
    \per
}
This observation suggests that $P_\mathrm{therm} \propto \gamma_w^1$.  
However, $\langle \Delta p_z \rangle$ may also increase or decrease with some power of $\gamma_w$.  
In the SCR formalism, the average longitudinal momentum transfer (due to single photon emission) may be calculated as 
\ba{\label{eq:avg_Delta_pz}
    \langle \Delta p_z \rangle = \int \! \dd \Pbb_{0 \to 0\gamma} \, \Delta p_z 
    \com
}
where $\dd \Pbb_{0\to 0\gamma}$ is the differential probability to emit a photon with momentum between $\pvec$ and $\pvec + \dd \pvec$, and where $\Delta p_z$ is the longitudinal momentum exchanged between the particles and the wall.
We take the longitudinal momentum transfer to be~\cite{Bodeker:2009qy} 
\bes{\label{eq:Delta_pz}
    \Delta p_z 
    & = p_{a,z} - p_{b,z} - p_z \\ 
    & = \bigl( E_a^2 - |\pvec_{a,\perp}|^2 - m_a^2 \bigr)^{1/2} - \bigl( E_b^2 - |\pvec_{b,\perp}|^2 - m_b^2 \bigr)^{1/2} - \bigl( E_p^2 - |\pvec_\perp|^2 \bigr)^{1/2} \\ 
    & \approx \bigl( E_a - E_b - E_p \bigr) - \frac{|\pvec_{a,\perp}|^2 + m_a^2}{2 E_a} + \frac{|\pvec_{b,\perp}|^2 + m_b^2}{2 E_b} + \frac{|\pvec_\perp|^2}{2 E_p} \\ 
    & \approx - \frac{|\pvec_{a,\perp}|^2 + m_a^2}{2 E_a} + \frac{|\pvec_{a,\perp} - \pvec_\perp|^2 + m_b^2}{2 (E_a - E_p)} + \frac{|\pvec_\perp|^2}{2 E_p} 
    \per
}
The transition from the first to the second line is achieved by assuming on-shell energy-momentum relations for all three particles. The second to the third line is based on the assumption \( E \approx p_z \gg |\pvec_\perp|, m \) for all three particles. 
Finally, moving from the third to the fourth line involves the assumption of energy and transverse momentum conservation among the three particles. 
In the subsequent sections, we will discuss the methodology for calculating \( \dd \Pbb_{0 \to 0\gamma} \) and provide evaluations using several simplified models for the current.

%==================================
% Vacuum persistence probability
%==================================
\section{Vacuum persistence probability}
\label{sec:vacuum}

%=========
In this section, we define the vacuum persistence probability and show how it can be calculated from the Fourier transform of the electromagnetic current density.  
The vacuum persistence probability is a quantity of interest, because unitarity links it to the total radiation probability.  
In other words, $1 = \Pbb_{0\to0} + \Pbb_{0\to\mathrm{any}}$ where the vacuum persistence probability (denoted by $\Pbb_{0\to0}$) quantifies the rate of zero photon emission and the total radiation probability (denoted by $\Pbb_{0\to\mathrm{any}}$) quantifies the rate of emitting any number of photons.  
The latter quantity is used to calculate the average longitudinal momentum transfer through \eref{eq:Delta_pz}.  

%-------------------------------------------
% Define amplitude and probability
%-------------------------------------------
\subsection{Define amplitude and probability}
\label{sub:vac_prob_def}

%=========
We define the vacuum persistence amplitude as follows.  
Since the theory has a time-dependent Hamiltonian (because of the source), there is no unique ground state.  
Instead, we distinguish two states called the $\IN$-vacuum $\ket{0_\IN}$ and the $\OUT$-vacuum $\ket{0_\OUT}$.  
The vacuum persistence amplitude is 
\ba{\label{eq:W00_def}
	W_{0\to0} = \amp{0_\OUT}{0_\IN} 
	\per
}
Note that $W_{0\to0}$ is simply a complex number.  
The squared modulus 
\ba{\label{eq:P00_def}
	\Pbb_{0\to0} = |W_{0\to0}|^2\,,
}
gives the vacuum persistence probability, also called the vacuum survival probability.  

%-------------------------------------------
% Calculate using the Gell-Mann and Low theorem
%-------------------------------------------
\subsection{Calculate using the Gell-Mann and Low theorem}
\label{sub:vac_prob_GML}

%=========
We calculate the vacuum persistence amplitude as follows.\footnote{Compare with Problem 4.1 of \rref{Peskin:1995ev}, which explores the vacuum survival probability for a scalar theory with a source, $\hat{H}_\mathrm{int}(t) = - \int \! \dd^3 \xvec \, j(x) \hat{\phi}(x)$.  }  
Using the Gell-Mann and Low theorem~\cite{Peskin:1995ev}, the $\IN$ and $\OUT$ vacuum states are written as a time evolution operator applied to the $\FREE$ vacuum state $\ket{0_\FREE}$, and the vacuum persistence amplitude can be expressed as
\ba{
	W_{0\to0} = \lim_{T\to\infty(1-\ii0)} \expval{0_\FREE}{\mathrm{Texp}\bigg\{ - \ii \int_{-T}^{T} \! \dd t \, \hat{H}_\mathrm{int}(t) \biggr\}}{0_\FREE} 
	\com
}
where $\mathrm{Texp}\{\cdot\}$ denotes the time-ordered operator exponential.   
Using the expression for $\hat{H}_\mathrm{int}(t)$ from \eref{eq:Hint} lets us write
\ba{\label{eq:W00_pre_Dyson}
	W_{0\to0} = \lim_{T\to\infty(1-\ii0)} \expval{0_\FREE}{\mathrm{Texp}\bigg\{ - \ii \int_T \! \dd^4 x \, j^\mu(x) \, \hat{A}_\mu(x) \biggr\}}{0_\FREE} 
	\com
}
where $\int_T \dd^4 x \equiv \int_{-T}^T \! \dd t \int \! \dd^3 \xvec$.  
Notice that the long-time limit must be performed with a small, negative imaginary part $-\ii0$ that acts as an IR regulator.  

%-------------------------------------------
% Employ perturbation theory
%-------------------------------------------
\subsection{Employ perturbation theory}
\label{sub:perturbation}

%=========
We use perturbation theory to evaluate $W_{0\to0}$.  
We assume that this quantity can be expressed as an infinite series in nonnegative powers of the electromagnetic coupling $e$, remembering that $j^\mu(x) \propto e$.  
If this is true, then the vacuum persistence amplitude can be written as 
\ba{\label{eq:W00_series}
	W_{0\to0} = \sum_{n=0}^\infty \frac{1}{n!} \, W_{0\to0}^{(n)} 
	\com
}
where $W_{0\to0}^{(n)} = O(e^n)$.  
By writing the exponential in \eref{eq:W00_pre_Dyson} in terms of its Taylor series expansion, we can read off the terms in the series.  
For the $n^\mathrm{th}$ term, we find
\bes{\label{eq:W00n}
	W_{0\to0}^{(n)} 
	& = \lim_{T\to\infty(1-\ii0)} 
	(-\ii)^n \int_T \! \dd^4 x_1 \cdots \int_T \! \dd^4 x_n \ 
	j^{\mu_1}(x_1) \cdots j^{\mu_n}(x_n) 
	\\ & \hspace{4cm} 
	\times \expval{0_\FREE}{\mathrm{T}\Big\{ \hat{A}_{\mu_1}(x_1) \cdots \hat{A}_{\mu_n}(x_n) \Bigr\}}{0_\FREE}
	\per
}

%-------------------------------------------
% Calculate $W_{0\to0}^{(n)}$ and sum the series 
%-------------------------------------------
\subsection{Calculate $W_{0\to0}^{(n)}$ and sum the series}
\label{sub:W00n}

%=========
Since the vacuum expectation value of an odd number of field operators vanishes, it follows that $W_{0\to0}^{(n)} = 0$ for odd $n$.  
Consequently the terms with odd powers of $e$ drop out of the series.  
To evaluate $W_{0\to0}^{(n)}$ for even $n$, one can use Wick's theorem to `factor' the $n$-point correlation function into products of two-point correlation functions.  
After accounting for the combinatorical factors, one finds\footnote{For this derivation, it is important that each field operator is evaluated at a distinct spacetime point.  In this sense, the field operators are all interchangeable, and everything boils down to copies of the same two-point function.  The derivation would have been different in an interacting field theory with multiple field operators evaluated at the same spacetime point.  This would lead to contractions of a point with itself, corresponding to Feynman graphs with loops.}
\ba{\label{eq:W00_even}
	\frac{1}{n!} \, W_{0\to0}^{(n)} 
	& = \frac{1}{(n/2)!} \biggl( \frac{W_{0\to0}^{(2)}}{2} \biggr)^{n/2} 
	\quad \text{for even $n$} 
    \per
}
Summing the series of odd and even $n$ terms leads to 
\ba{\label{eq:W00_from_W002}
	W_{0\to0} 
	= \mathrm{exp}\biggl( \frac{W_{0\to0}^{(2)}}{2} \biggr) 
	\per
}
Using \eref{eq:P00_def} the vacuum survival probability is found to be
\ba{\label{eq:P00_from_W002}
	\Pbb_{0\to0} 
    = \mathrm{exp}\Bigl( \mathrm{Re} \, W_{0\to0}^{(2)} \Bigr) 
	\per
}
It must be the case that $\mathrm{Re} \, W_{0\to0}^{(2)} \leq 0$ to ensure $0 \leq \Pbb_{0\to0} \leq 1$. 

%-------------------------------------------
% Derive expression for $W_{0\to0}^{(2)}$
%-------------------------------------------
\subsection{Derive expression for $W_{0\to0}^{(2)}$}

%=========
The $O(e^2)$ contribution to the vacuum survival amplitude is given by 
\ba{\label{eq:W002_def}
	W_{0\to0}^{(2)} 
	& = \lim_{T\to\infty(1-\ii0)} 
	(-1) \int_T \! \dd^4 x 
	\int_T \! \dd^4 y \ 
	j^{\mu}(x) \, j^{\nu}(y) \, 
	\expval{0_\FREE}{\mathrm{T}\Bigl\{ \hat{A}_{\mu}(x) \, \hat{A}_{\nu}(y) \Bigr\} }{0_\FREE}
	\per 
}
The source at point $y^\mu$ (or $x^\mu$ if it is earlier) creates a photon that propagates to $x^\mu$ (or $y^\mu$ if it is later) where it is absorbed by the source.  
The last factor is the photon Feynman propagator $D_{F,\mu\nu}(x-y)$ from \eref{eq:DF_def}.  
If we write the Feynman propagator using its Fourier representation from \eref{eq:DF_Fourier}, and if we ignore the $-\ii0$ shifts in the limits of integration, then the $x^\mu$ and $y^\mu$ integrals simply yield the Fourier transform of the current; see \eref{eq:Fourier_transform}.  
Making this replacement gives
\ba{\label{eq:W002_from_jj}
	W_{0\to0}^{(2)} 
%---
	= \lim_{\epsF \to 0^+} \int \! \! \frac{\dd^4 k}{(2\pi)^4} \, \frac{\ii}{k \cdot k + \ii \epsF} \, \tilde{j}(k)^\ast \cdot \tilde{j}(k) 
	\per
}
The vacuum survival probability is then calculated using \eref{eq:P00_from_W002}. 
In the later sections we perform these calculations using simple models for the particle worldline.  

%==================================
% Single emission probability
%==================================
\section{Single emission probability}
\label{sec:emission}

%=========
In this section, we define the single photon emission probability and show how it can be calculated from the Fourier transform of the electromagnetic current density.  
Although the vacuum survival probability $\Pbb_{0\to0}$ can be used to infer the total emission probability $\Pbb_{0\to\mathrm{any}} = 1 - \Pbb_{0\to0}$, it does not provide information about the spectrum of emission.  
By calculating the single photon emission probability, we also determine the energy spectrum of the radiation.  
This information is needed to calculate the average longitudinal momentum transfer $\langle \Delta p_z \rangle$ using \eref{eq:Delta_pz}.  

%-------------------------------------------
% Define amplitude and probability
%-------------------------------------------
\subsection{Define amplitude and probability}
\label{sec:emission_def}

%=========
We define the single photon emission amplitude as 
\ba{\label{eq:W00g_def}
	W_{0\to0\gamma}(\pvec,s) = \amp{(\pvec, s)_\OUT}{0_\IN} 
	\com
}
where $\ket{0_\IN}$ is the $\IN$-vacuum and where $\ket{(\pvec, s)_\OUT}$ is a one-particle state built upon the $\OUT$-vacuum in which the photon has three-momentum $\pvec$ and spin projection $s$.  
The differential probability for emitting a photon with momentum between $\pvec$ and $\pvec + \dd \pvec$ is given by 
\ba{\label{eq:dP00g_def}
	\dd\Pbb_{0\to0\gamma}(\pvec)
	= \frac{\dd^3 \pvec}{(2\pi)^3} \, \frac{1}{2E_p} \, \sum_{s = \pm 1} \bigl| W_{0\to0\gamma} \bigr|^2 
	\com
}
where $E_p = |\pvec|$ for a massless photon.  
Note that we sum over $s$, since we are not interested in the helicity of the outgoing photon.  
Integrating over the full phase space for the emitted photon gives 
\ba{\label{eq:P00g_def}
	\Pbb_{0\to0\gamma} 
	= \int \! \dd \Pbb_{0\to0\gamma} 
	= \int \! \! \frac{\dd^3 \pvec}{(2\pi)^3} \, \frac{1}{2E_p} \, \sum_{s = \pm 1} \bigl| W_{0\to0\gamma} \bigr|^2 
	\com
}
which is the single emission probability.  

%-------------------------------------------
% Calculate using the Gell-Mann and Low theorem
%-------------------------------------------
\subsection{Calculate using the Gell-Mann and Low theorem}
\label{sec:emission_GML}

%=========
We calculate the matrix element using the Gell-Mann and Low theorem: 
\bes{
	W_{0\to0\gamma}(\pvec,s) 
	& = \lim_{T\to\infty(1-\ii0)} \, \expval{(\pvec, s)_\FREE}{\mathrm{Texp}\bigg\{ - \ii \int_T \! \dd^4 x \, j^\mu(x) \, \hat{A}_\mu(x) \biggr\}}{0_\FREE} 
	\com
}
where the free $\FREE$ one-particle state is defined in \eref{eq:ps_FREE}.  

%-------------------------------------------
% Employ perturbation theory
%-------------------------------------------
\subsection{Employ perturbation theory}
\label{sec:emission_perturbation}

%=========
We employ perturbation theory to evaluate the matrix element order-by-order in powers of $e$.  
We write the series as 
\ba{\label{eq:W00g_series}
	W_{0\to0\gamma}(\pvec,s) & = \sum_{n=0}^{\infty} \frac{1}{n!} \, W_{0\to0\gamma}^{(n)}(\pvec,s) 
	\com
}
where $W_{0\to0\gamma}^{(n)} = O(e^n)$.  
The $n^\mathrm{th}$ term is given by 
\bes{\label{eq:W00gn_def}
	W_{0\to0\gamma}^{(n)}(\pvec,s) 
	& = (-\ii)^n \, \lim_{T\to\infty(1-\ii0)} 
	\int_T \! \dd^4 x_1 \cdots \dd^4 x_n \ 
	j^{\mu_1}(x_1) \cdots j^{\mu_n}(x_n) 
	\\ & \hspace{4cm} 
	\times \expval{(\pvec,s)_\FREE}{\mathrm{T}\Big\{ \hat{A}_{\mu_1}(x_1) \cdots \hat{A}_{\mu_n}(x_n) \Bigr\}}{0_\FREE}
	\per
}
In particular the $O(e^1)$ term takes the form,
\ba{
	W_{0\to0\gamma}^{(1)}(\pvec,s) 
	& = \lim_{T\to\infty(1-\ii0)} 
	(-\ii) \, \int_T \! \dd^4 x \ 
	j^{\mu}(x) \ 
	\expval{(\pvec,s)_\FREE}{ \, \hat{A}_{\mu}(x) \, }{0_\FREE} 
	\per
}
The matrix element factor is simply the mode function for a photon with three-momentum $\pvec$ and spin projection $s$, which is written as 
\ba{\label{eq:W00g1_def}
	W_{0\to0\gamma}^{(1)}(\pvec,s) 
	& = \lim_{T\to\infty(1-\ii0)} 
	(-\ii) \, \int_T \! \dd^4 x \ 
	j(x) \cdot \varepsilon^\ast(\pvec,s) \, \ee^{\ii p \cdot x} 
	\com
}
where $p^\mu = (E_p, \pvec)$ and $E_p = |\pvec|$.  

%-------------------------------------------
% Calculate $W_{0\to0\gamma}^{(n)}$ and sum the series
%-------------------------------------------
\subsection{Calculate $W_{0\to0\gamma}^{(n)}$ and sum the series}
\label{sec:emission_sum}

%=========
Since an even number of field operators cannot connect the $\FREE$ vacuum state with a $\FREE$ one-particle state, it is clear that $W_{0\to0\gamma}^{(n)} = 0$ for even $n$.  
For the terms with odd $n$ one can derive the relation
\bes{
	W_{0\to0\gamma}^{(2n+1)} & = \bigl( 2n + 1 \bigr) \, W_{0\to0\gamma}^{(1)} \, W_{0\to0}^{(2n)} 
	\com
}
where $W_{0\to0}^{(2n)}$ is the $O(e^{2n})$ term in the vacuum persistence amplitude given by \eref{eq:W00_even}, and an expression for $W_{0\to0\gamma}^{(1)}$ appears in \eref{eq:W00g1_def}.  
Using \eref{eq:W00_even} lets us write this relation as 
\ba{
	\frac{W_{0\to0\gamma}^{(2n+1)}}{(2n+1)!} & = W_{0\to0\gamma}^{(1)} \, \frac{1}{n!} \biggl( \frac{W_{0\to0}^{(2)}}{2} \biggr)^n 
	\per
}
Upon summing over the even and odd $n$ terms, the factors of $W_{0\to0}^{(2)}$ exponentiate into the vacuum persistence amplitude $W_{0\to0} = \mathrm{exp}(W_{0\to0}^{(2)} / 2)$, and the single emission amplitude is given by 
\ba{\label{eq:W00g_from_W00g1}
	W_{0\to0\gamma} 
	= W_{0\to0} \times W_{0\to0\gamma}^{(1)} 
    \per
}
Putting this expression into the single emission probability from \eref{eq:dP00g_def} leads to 
\ba{\label{eq:dP00g_from_W00g1}
	\dd\Pbb_{0\to0\gamma}
	= \Pbb_{0\to0} \times \frac{\dd^3 \pvec}{(2\pi)^3} \, \frac{1}{2E_p} \, \sum_{s = \pm 1} \bigl| W_{0\to0\gamma}^{(1)} \bigr|^2 
	\com
}
where we have also used $\Pbb_{0\to0} = |W_{0\to0}|^2$ from \eref{eq:P00_from_W002}. 
The probability for emitting a single photon is simply the product of the vacuum survival probability (no photon emission) and the squared modulus of the leading $O(e)$ amplitude for photon emission. 

%-------------------------------------------
% Derive expression for $W_{0\to0\gamma}^{(1)}$
%-------------------------------------------
\subsection{Derive expression for $W_{0\to0\gamma}^{(1)}$}
\label{sec:emission_W00g1}

%=========
To evaluate $W_{0\to0\gamma}^{(1)}$ using \eref{eq:W00g1_def}, we drop the IR regulator $-\ii0$.  
Then resultant integral is simply the Fourier transform of the current density from \eref{eq:Fourier_transform}, which gives 
\bes{
	W_{0\to0\gamma}^{(1)}(\pvec,s) 
	& = 
	(-\ii) \, \tilde{j}(p) \cdot \varepsilon^\ast(\pvec,s) 
	\per
} 
Then the single emission probability from \eref{eq:dP00g_from_W00g1} takes the form
\bes{
	\dd\Pbb_{0\to0\gamma}
	& = \Pbb_{0\to0} \times \frac{\dd^3 \pvec}{(2\pi)^3} \, \frac{1}{2E_p} \, \tilde{j}^\mu(p)^\ast \tilde{j}^\nu(p) \sum_{s = \pm 1} \varepsilon_\mu(\pvec,s) \varepsilon_\nu^\ast(\pvec,s)
	\per
}
The sum over polarization vectors is simply $C_{\mu\nu}(\pvec)$ from \eref{eq:C_def}.  
We work in the Feynman gauge with $C_{\mu\nu}(\pvec) = - g_{\mu\nu}$.  
This leads to the final expression for the differential photon emission probability
\ba{\label{eq:dP00g_from_jj}
	\dd\Pbb_{0\to0\gamma}
	= - \Pbb_{0\to0} \times \frac{\dd^3 \pvec}{(2\pi)^3} \, \frac{1}{2E_p} \, \tilde{j}(p)^\ast \cdot \tilde{j}(p) 
	\per
}
In order for the differential probability to be nonnegative $\dd\Pbb_{0\to0\gamma} \geq 0$, we require $\tilde{j}(p)^\ast \cdot \tilde{j}(p) \leq 0$.  

%==================================
% Model 1:~~Constant velocity
%==================================
\section{Model 1:~~Constant velocity}
\label{sec:model1}

%=========
As a starting point, suppose that the charged particle travels with a constant velocity.  
Since it does not accelerate, we expect to find that there is no radiation. 
We confirm this expectation by calculating the vacuum persistence probability (showing that it equals $1$) and the single emission probability (showing that it equals $0$).

%-------------------------------------------
% Current density
%-------------------------------------------
\subsection{Current density}
\label{sub:model1_current}

%=========
If the particle has a constant velocity $\vvec(\tau) = \vvec$ in some frame of reference then the worldline, four-velocity, and four-acceleration are 
\bes{\label{eq:model1_worldline}
    \Xbb^\mu(\tau) = \Xbb_0^\mu + ( \tau - \tau_0 ) \, \Ubb^\mu 
    \ , \qquad 
    \Ubb^\mu = \bigl( \gamma \, , \, \gamma \vvec \bigr) 
    \ , \qquad \text{and} \qquad 
    \Abb^\mu & = \bigl( 0 \, , \, {\bm 0} \bigr) \,,
}
where $\gamma = 1 / \sqrt{1 - |\vvec|^2}$ is the Lorentz factor.  
We impose the initial condition $\Xbb^\mu(\tau) = \Xbb_0^\mu$ at $\tau = \tau_0$.  
The Fourier transform of the current density is calculated by evaluating the integral in \eref{eq:jmuFT}.  
Doing so gives 
\bes{\label{eq:model1_current}
	\tilde{j}^\mu(k) 
	& = \qe \int_{-\infty}^{\infty} \! \dd\tau \, \Ubb^\mu(\tau) \, \ee^{\ii k \cdot \Xbb(\tau)} \\ 
	& = \qe \int_{-\infty}^{\infty} \! \dd\tau \, \Ubb^\mu \, \ee^{\ii k \cdot (\Xbb_0 + \Ubb \tau - \Ubb \tau_0)} \\ 
    & = \qe \, \Ubb^\mu \, \ee^{\ii k \cdot \Xbb_0} \int_{-\infty}^{\infty} \! \dd\tau \, \ee^{\ii k \cdot \Ubb (\tau - \tau_0)} \\ 
	& = \qe \, \Ubb^\mu \, \ee^{\ii k \cdot \Xbb_0} \, (2\pi) \delta(k \cdot \Ubb) 
	\com
}
with the Dirac delta factor  written as 
\ba{
	\delta(k \cdot \Ubb) = \gamma^{-1} \, \delta(\omega - \kvec \cdot \vvec) 
	\com
}
where we have written the wavevector as $k^\mu = (\omega, \kvec)$.  
This factor vanishes for wavevectors that do not obey $\omega = \kvec \cdot \vvec$.  
For instance, if the particle is at rest ($\vvec = 0$) then the current density selects out the static modes ($\omega = 0$).  
Boosting to a frame where the particle is in (constant) motion gives $\omega = \kvec \cdot \vvec$ instead.  
Also note that $k \cdot \tilde{j}(k) = 0$ as required by the continuity equation.  

%-------------------------------------------
% Vacuum persistence
%-------------------------------------------
\subsection{Vacuum persistence}
\label{sub:model1_vacuum}

%=========
Expressions for the vacuum persistence amplitude and the vacuum survival probability were provided in \erefs{eq:W00_from_W002}{eq:P00_from_W002}:  $W_{0\to0} = \mathrm{exp}( W_{0\to0}^{(2)} / 2)$ and $\Pbb_{0\to0} = \mathrm{exp}(\mathrm{Re} \, W_{0\to0}^{(2)})$.   
Then \eref{eq:W002_from_jj} expresses $W_{0\to0}^{(2)}$ in terms of the current density $\tilde{j}(k)$.  
Using the expression for $\tilde{j}(k)$ from \eref{eq:model1_current} leads to 
\bes{
	W_{0\to0}^{(2)} 
	& = \lim_{\epsF \to 0^+} \, (\qe)^2 \int \! \! \frac{\dd^4 k}{(2\pi)^4} \, \frac{\ii}{k \cdot k + \ii \epsF} \, (2\pi)^2 \delta(k \cdot \Ubb) \, \delta(k \cdot \Ubb) 
	\com
}
where we have also used $\Ubb \cdot \Ubb = 1$ from \eref{eq:U_norm}.  
Writing the wavevector as $k^\mu = (\omega, \kvec)$ leads to 
\ba{
	W_{0\to0}^{(2)} 
	= \lim_{\epsF \to 0^+} \, \frac{(\qe)^2}{\gamma^2} \int \! \frac{\dd \omega}{2\pi} \frac{\dd^3 \kvec}{(2\pi)^3} \, \frac{\ii}{\omega^2 - |\kvec|^2 + \ii \epsF} \, (2\pi)^2 \delta(\omega - \kvec \cdot \vvec) \, \delta(\omega - \kvec \cdot \vvec) 
	\per
}
One of the Dirac delta functions may be used to evaluate the integral over $\omega$, which leads to 
\bes{
	W_{0\to0}^{(2)} 
	& = \lim_{\epsF \to 0^+} \, \frac{(\qe)^2}{\gamma^2} \, (2\pi) \delta(0) \int \! \! \frac{\dd^3 \kvec}{(2\pi)^3} \, \frac{\ii}{|\kvec \cdot \vvec|^2 - |\kvec|^2 + \ii \epsF} 
	\per
}
The singular factor $(2\pi) \delta(0)$ has units of time.  
If we had only allowed the particle to exist for a finite time interval $T$, then this factor would be replaced by $(2\pi) \delta(0) \to T$.  
For large $T$, the amplitude is enhanced by the long duration of the source.  
To identify a finite quantity, we can interpret $W_{0\to0}^{(2)} / (2\pi)\delta(0)$ as the vacuum survival rate. 

%=========
To evaluate the integral over $\kvec$ we work in spherical coordinates where $\vvec$ defines the polar axis:  $k = |\kvec|$ is the wavenumber radial coordinate, $\theta$ is the polar angular coordinate, and $\phi$ is the azimuthal angular coordinate.  
The integral over $\phi$ is trivial, providing a factor of $2\pi$.  
The remaining integrals are 
\ba{
	W_{0\to0}^{(2)} 
	& = \Bigl( -\ii (\qe)^2 \, (2\pi) \delta(0) \Bigr) \biggl( \int_0^\infty \! \frac{\dd k}{2\pi^2} \biggr) \biggl( \lim_{\epsF \to 0^+} \, \frac{1}{\gamma^2} \int_{-1}^1 \! \frac{\dd \cos\theta}{2} \, \frac{-1}{v^2 \cos^2 \theta - 1 + \ii \epsF / k^2} \biggr) 
	\com
}
where $v = |\vvec|$.  
For $0 \leq v < 1$ there is no IR divergence within the integration domain, and the IR regulator $\epsF$ can be safely removed.  
The integral over $\cos\theta$ evaluates to 
\bes{
	\frac{1}{\gamma^2} \int_{-1}^1 \! \frac{\dd \cos\theta}{2} \, \frac{-1}{v^2 \cos^2 \theta - 1} 
	& = \frac{1-v^2}{2v} \log \frac{1-v}{1+v} 
	\per
}
This factor approaches $1$ as $v \to 0$ from above, and it approaches $0$ as $v \to 1$ from below.  
The integral over $k$ has a UV divergence, which we can regulate by replacing the upper limit of integration with a UV cutoff $k_\UV$.  
Then the regulated integral evaluates to 
\ba{
	\lim_{k_\UV \to \infty} \int_0^{k_\UV} \! \frac{\dd k}{2\pi^2} 
	= \lim_{k_\UV \to \infty} \frac{k_\UV}{2\pi^2} 
	\per
}
Since the source is localized to a point in space, its Fourier transform has support at arbitrarily large $k$, corresponding to arbitrarily short length scales; this is the source of a UV divergence.  
Putting all these factors together gives 
\ba{
	W_{0\to0}^{(2)} 
	= \lim_{k_\UV \to \infty} -\ii (\qe)^2 \, (2\pi) \delta(0) \frac{k_\UV}{2\pi^2} \frac{1-v^2}{2v} \log \frac{1-v}{1+v} 
	\per
}
The vacuum survival probability $\Pbb_{0\to0}$ is then given by \eref{eq:P00_from_W002}.  
Since $W_{0\to0}^{(2)}$ is pure imaginary, with vanishing real part, it follows that 
\ba{\label{eq:model1_P00}
	\Pbb_{0\to0} = \mathrm{exp}\Bigl( \mathrm{Re} \, W_{0\to0}^{(2)} \Bigr) = 1 
	\com
}
and conversely $\Pbb_{0\to\mathrm{any}} = 1 - \Pbb_{0\to0} = 0$.  
Despite the dangerous divergences in $W_{0\to0}^{(2)}$, we do obtain the expected result.  
A charged particle with constant velocity has unit probability to emit zero photons, and zero probability to emit any number of photons.  

%-------------------------------------------
% Single emission
%-------------------------------------------
\subsection{Single emission}
\label{sub:model1_emission}

%=========
An expression for the differential probability of a single photon emission was provided in \eref{eq:dP00g_from_jj}. 
Using the expression for $\tilde{j}(k)$ from \eref{eq:model1_current} leads to 
\ba{
	\dd\Pbb_{0\to0\gamma} 
	& = - \frac{(\qe)^2}{\gamma^2} \frac{\dd^3 \pvec}{(2\pi)^3} \, \frac{1}{2E_p} \, \bigl[ (2\pi) \delta(E_p - \pvec \cdot \vvec) \bigr]^2 
    \com
}
where $E_p = |\pvec|$.  
In the derivation, we have used $\Ubb \cdot \Ubb = 1$ from \eref{eq:U_norm} and $\Pbb_{0\to0} = 1$ from \eref{eq:model1_P00}
If the particle travels with a speed $0 \leq v < 1$, \textit{i.e.} it is a massive particle, then the argument of the Dirac delta function cannot equal zero.  
Consequently, the single emission probability vanishes: 
\ba{
	\dd\Pbb_{0\to0\gamma} 
	& = 0 
	\per
}
This conclusion is consistent with our earlier result $\Pbb_{0\to0} = 1$: if the vacuum survival probability is one, the probability for any emission processes must be zero.  
Note that this conclusion would still hold even if the radiated particle were not a photon, but a massive particle instead with $E_p = (|\pvec|^2 + m^2)^{1/2} > |\pvec|$.  

%==================================
% Model 2:~~Abrupt deceleration
%==================================
\section{Model 2:~~Abrupt deceleration}
\label{sec:model2}

%=========
Next we consider a situation in which the charged particle abruptly decelerates at one instant in time.  
If the velocity and acceleration are collinear, it is customary to call this bremstrahlung radiation.  
In the classical theory of electromagnetism, bremstrahlung radiation is discussed in most textbooks including \rrefs{Jackson:1998nia,Griffiths:2024}.  
Whereas our analysis accounts for the quantum nature of the emitted photons, we find that the resultant radiation spectrum is equivalent to the familiar classical result. 
In this section and the next, we illuminate how the UV part of the emission spectrum depends sensitively on the abruptness of the transition, which is also a familiar behavior from classical electromagnetism~\cite{Jackson:1998nia}. 

%-------------------------------------------
% Current density
%-------------------------------------------
\subsection{Current density}
\label{sub:model2_current}

%=========
Suppose that the charged particle travels with velocity $\vvec_a$ and abruptly decelerates to velocity $\vvec_b$ at one instant in time.  
The four-velocity of the charged particle can be written as 
\ba{
	\Ubb^\mu(\tau) = \begin{cases}
	\Ubb_a^\mu = \bigl( \gamma_a \, , \, \gamma_a \vvec_a \bigr) & , \quad \tau < \tau_0 \\ 
	\Ubb_b^\mu = \bigl( \gamma_b \, , \, \gamma_b \vvec_b \bigr) & , \quad \tau > \tau_0 
	\end{cases}
	\com
}
where we assume that the acceleration happens at proper time $\tau = \tau_0$.  
The Lorentz factors are defined by $\gamma_{a,b} = 1 / \sqrt{1 - |\vvec_{a,b}|^2}$.  
Note that $\Ubb^\mu(\tau)$ is undefined at $\tau = \tau_0$.  
The corresponding worldline is 
\ba{\label{eq:model2_worldline}
	\Xbb^\mu(\tau) = \begin{cases}
	\Xbb_a^\mu(\tau) \equiv \Xbb_0^\mu + (\tau - \tau_0) \, \Ubb_a^\mu & , \quad \tau < \tau_0 \\ 
    \Xbb_0^\mu & , \quad \tau = \tau_0 \\ 
	\Xbb_b^\mu(\tau) \equiv \Xbb_0^\mu + (\tau - \tau_0) \, \Ubb_b^\mu & , \quad \tau > \tau_0 
	\end{cases}
	\com
}
and $\Xbb^\mu(\tau)$ is continuous at $\tau = \tau_0$.  
The Fourier transform of the current density is calculated by performing the proper time integral in \eref{eq:jmuFT}: 
\ba{\label{eq:model2_current}
	\tilde{j}^\mu(k) 
	& = \qe \int_{-\infty}^{\infty} \! \dd\tau \, \Ubb^\mu(\tau) \, \ee^{\ii k \cdot \Xbb(\tau)} \\ 
	& = \qe \int_{-\infty}^{\tau_0} \! \dd\tau \, \Ubb_a^\mu \, \ee^{\ii k \cdot (\Xbb_0 + \Ubb_a \tau - \Ubb_a \tau_0)} + \qe \int_{\tau_0}^{\infty} \! \dd\tau \, \Ubb_b^\mu \, \ee^{\ii k \cdot (\Xbb_0 + \Ubb_b \tau - \Ubb_b \tau_0)} \nn 
	& = \qe \, \ee^{\ii k \cdot \Xbb_0} \biggl[ \Ubb_a^\mu \, \ee^{- \ii k \cdot \Ubb_a \tau_0} \int_{-\infty}^{\tau_0} \! \dd\tau \, \ee^{\ii k \cdot \Ubb_a \tau} + \Ubb_b^\mu \, \ee^{-\ii k \cdot \Ubb_b \tau_0} \int_{\tau_0}^{\infty} \! \dd\tau \, \ee^{\ii k \cdot \Ubb_b \tau} \biggr] \nn 
	& = \lim_{\varepsilon \to 0^+} \qe \, \ee^{\ii k \cdot \Xbb_0} \biggl[ \Ubb_a^\mu \, \ee^{- \ii k \cdot \Ubb_a \tau_0} \int_{-\infty}^{\tau_0} \! \dd\tau \, \ee^{\ii (k \cdot \Ubb_a - \ii \varepsilon) \tau} + \Ubb_b^\mu \, \ee^{- \ii k \cdot \Ubb_b \tau_0} \int_{\tau_0}^{\infty} \! \dd\tau \, \ee^{\ii (k \cdot \Ubb_b + \ii \varepsilon) \tau} \biggr] \nn 
	& = \lim_{\varepsilon \to 0^+} (-\ii) \qe \, \ee^{\ii k \cdot \Xbb_0} \biggl[ \frac{\Ubb_a^\mu}{k \cdot \Ubb_a - \ii \varepsilon} - \frac{\Ubb_b^\mu}{k \cdot \Ubb_b + \ii \varepsilon} \biggr] 
	\per
    \nonumber
}
Here it was necessary to introduce a small positive parameter $\varepsilon$, which acts as an IR regulator,  which must be sent to zero at the end of the calculation.  
For $\Ubb_a = \Ubb_b$ we regain the current density calculated in \sref{sec:model1} for the constant-velocity model; \textit{cf.} \eref{eq:model1_current}.\footnote{The identity $1 / (z \pm \ii \varepsilon) = \mathrm{PV}(1/z) \mp \ii \pi \delta(z)$ implies $1 / (z - \ii \varepsilon) - 1 / (z + \ii \varepsilon) = 2 \pi \ii \delta(z)$.}  
It is also easy to see that $k \cdot \tilde{j}(k) = 0$ in the limit $\varepsilon \to 0^+$, which is required by the continuity equation.  

%=========
Whereas this calculation can be performed for general $\vvec_a$ and $\vvec_b$, the angular integrals simplify if we assume that these vectors are parallel.  
In practice, we will occasionally assume 
\ba{\label{eq:model2_assume}
    \vvec_a \parallel \vvec_b 
    \ , \qquad 
    \vvec_a = v_a \, \vhat 
    \ , \qquad 
    \vvec_b = v_b \, \vhat 
    \ , \qquad \text{and} \qquad 
    0 \leq v_b < v_a < 1 
    \com
}
where $\vhat$ is a unit vector.  
Under these assumptions, the motion of the charged particle remains in a straight line and experiences a sudden deceleration but continues moving forward.  
This is the familiar setup for a calculation of bremstrahlung radiation.  
This is a reasonable approximation for the motion of an ultrarelativistic particle near a Higgs-phase bubble wall where the particle's mass increases across the bubble wall interface.  
In the rest frame of the wall, the particle is highly boosted, and its velocity is approximately normal to the plane of the wall.  

%-------------------------------------------
% Vacuum persistence
%-------------------------------------------
\subsection{Vacuum persistence}
\label{sub:model2_vacuum}

%=========
Expressions for the vacuum persistence amplitude and the vacuum survival probability were provided in \erefs{eq:W00_from_W002}{eq:P00_from_W002}:  $W_{0\to0} = \mathrm{exp}( W_{0\to0}^{(2)} / 2)$ and $\Pbb_{0\to0} = \mathrm{exp}(\mathrm{Re} \, W_{0\to0}^{(2)})$, respectively.   
Then \eref{eq:W002_from_jj} expresses $W_{0\to0}^{(2)}$ in terms of the current density $\tilde{j}(k)$.  
Using the expression for $\tilde{j}(k)$ from \eref{eq:model2_current}, this leads to
\ba{
	W_{0\to0}^{(2)} 
	& = \lim_{\epsF, \varepsilon \to 0^+} (\qe)^2 \int \! \! \frac{\dd^4 k}{(2\pi)^4} \, \frac{\ii}{k \cdot k + \ii \epsF} \, \biggl[ 
	\frac{\Ubb_a \cdot \Ubb_a}{\bigl( k \cdot \Ubb_a + \ii \varepsilon \bigr) \bigl( k \cdot \Ubb_a - \ii \varepsilon \bigr)} 
	- \frac{\Ubb_a \cdot \Ubb_b}{\bigl( k \cdot \Ubb_a + \ii \varepsilon \bigr) \bigl( k \cdot \Ubb_b + \ii \varepsilon \bigr)} 
	\nn & \hspace{4cm}
	- \frac{\Ubb_b \cdot \Ubb_a}{\bigl( k \cdot \Ubb_b - \ii \varepsilon \bigr) \bigl( k \cdot \Ubb_a - \ii \varepsilon \bigr)} 
	+ \frac{\Ubb_b \cdot \Ubb_b}{\bigl( k \cdot \Ubb_b - \ii \varepsilon \bigr) \bigl( k \cdot \Ubb_b + \ii \varepsilon \bigr)} \biggr] 
	\per 
}
Writing out the four-vectors $k^\mu = (\omega, \kvec)$, $\Ubb_a^\mu = (\gamma_a, \gamma_a \vvec_a)$, and $\Ubb_b^\mu = (\gamma_b, \gamma_b \vvec_b)$ leads to 
\ba{\label{eq:model2_W002_from_I}
	W_{0\to0}^{(2)} 
	& = (\qe)^2 \bigl( \Ical_{aa} - \Ical_{ab} - \Ical_{ba} + \Ical_{bb} \bigr) 
	\com
}
where we have defined 
\bsa{}{
%---
	\Ical_{aa} & = \lim_{\epsF, \varepsilon \to 0^+} \ii \int \! \frac{\dd \omega}{2\pi} \int \! \! \frac{\dd^3 \kvec}{(2\pi)^3} \, \frac{1 - \vvec_a \cdot \vvec_a}{\bigl( \omega - |\kvec| + \ii \epsF \bigr) \bigl( \omega + |\kvec| - \ii \epsF \bigr) \bigl( \omega - \kvec \cdot \vvec_a + \ii \varepsilon \bigr) \bigl( \omega - \kvec \cdot \vvec_a - \ii \varepsilon \bigr)}\,, \\ 
%---
	\Ical_{ab} & = \lim_{\epsF, \varepsilon \to 0^+} \ii \int \! \frac{\dd \omega}{2\pi} \int \! \! \frac{\dd^3 \kvec}{(2\pi)^3} \, \frac{1 - \vvec_a \cdot \vvec_b}{\bigl( \omega - |\kvec| + \ii \epsF \bigr) \bigl( \omega + |\kvec| - \ii \epsF \bigr) \bigl( \omega - \kvec \cdot \vvec_a + \ii \varepsilon \bigr) \bigl( \omega - \kvec \cdot \vvec_b + \ii \varepsilon \bigr)}\,, \\ 
%---
	\Ical_{ba} & = \lim_{\epsF, \varepsilon \to 0^+} \ii \int \! \frac{\dd \omega}{2\pi} \int \! \! \frac{\dd^3 \kvec}{(2\pi)^3} \, \frac{1 - \vvec_b \cdot \vvec_a}{\bigl( \omega - |\kvec| + \ii \epsF \bigr) \bigl( \omega + |\kvec| - \ii \epsF \bigr) \bigl( \omega - \kvec \cdot \vvec_b - \ii \varepsilon \bigr) \bigl( \omega - \kvec \cdot \vvec_a - \ii \varepsilon \bigr)}\,, \\ 
%---
	\Ical_{bb} & = \lim_{\epsF, \varepsilon \to 0^+} \ii \int \! \frac{\dd \omega}{2\pi} \int \! \! \frac{\dd^3 \kvec}{(2\pi)^3} \, \frac{1 - \vvec_b \cdot \vvec_b}{\bigl( \omega - |\kvec| + \ii \epsF \bigr) \bigl( \omega + |\kvec| - \ii \epsF \bigr) \bigl( \omega - \kvec \cdot \vvec_b - \ii \varepsilon \bigr) \bigl( \omega - \kvec \cdot \vvec_b + \ii \varepsilon \bigr)} 
	\per
}
Next, we evaluate the $\omega$ integrals by analytically continuing the integration path into the complex $\omega$-plane.  
Each integrand has four poles.  
Integrals $\Ical_{aa}$ and $\Ical_{bb}$ both have two poles in the upper half plane and two poles in the lower half plane.  
Integral $\Ical_{ba}$ only has one pole in the lower half plane, and integral $\Ical_{ab}$ only has one pole in the upper half-plane.  
We close the integration contour in either the upper or lower half-plane, depending on which has fewer poles.  
Closing in the lower half plane brings a factor of $-2\pi\ii$ times the residue, and closing in the upper half plane brings a factor of $+2\pi\ii$ times the residue.  
The integral along the arc off the real axis vanishes since $\dd \omega / \omega^4 \to 0$ for large $|\omega|$.  
Evaluating the $\omega$ integrals in this way gives 
\bsa{}{
%---
	\Ical_{aa} & = \lim_{\epsF, \varepsilon \to 0^+} \frac{1}{2} \int \! \! \frac{\dd^3 \kvec}{(2\pi)^3} \, \biggl( 
	\frac{1 - \vvec_a \cdot \vvec_a}{\bigl( |\kvec| - \ii \epsF \bigr) \bigl( |\kvec| + \kvec \cdot \vvec_a - \ii \epsP \bigr) \bigl( |\kvec| + \kvec \cdot \vvec_a - \ii \epsM \bigr)} 
	\\ & \hspace{4cm}
	+ \frac{1 - \vvec_a \cdot \vvec_a}{\bigl( |\kvec| - \kvec \cdot \vvec_a - \ii \epsP \bigr) \bigl( |\kvec| + \kvec \cdot \vvec_a - \ii \epsM \bigr) \bigl( \ii \varepsilon \bigr)}  
	\biggr)\,, \nn 
%---
	\Ical_{ab} & = \lim_{\epsF, \varepsilon \to 0^+} \frac{1}{2} \int \! \! \frac{\dd^3 \kvec}{(2\pi)^3} \, \frac{1 - \vvec_a \cdot \vvec_b}{\bigl( |\kvec| - \ii \epsF \bigr) \bigl( |\kvec| + \kvec \cdot \vvec_a - \ii \epsP \bigr) \bigl( |\kvec| + \kvec \cdot \vvec_b - \ii \epsP \bigr)} \,,\\ 
%---
	\Ical_{ba} & = \lim_{\epsF, \varepsilon \to 0^+} \frac{1}{2} \int \! \! \frac{\dd^3 \kvec}{(2\pi)^3} \, \frac{1 - \vvec_b \cdot \vvec_a}{\bigl( |\kvec| - \ii \epsF \bigr) \bigl( |\kvec| - \kvec \cdot \vvec_b - \ii \epsP \bigr) \bigl( |\kvec| - \kvec \cdot \vvec_a - \ii \epsP \bigr)} \,,\\ 
%---
	\Ical_{bb} & = \lim_{\epsF, \varepsilon \to 0^+} \frac{1}{2} \int \! \! \frac{\dd^3 \kvec}{(2\pi)^3} \, \biggl( 
	\frac{1 - \vvec_b \cdot \vvec_b}{\bigl( |\kvec| - \ii \epsF \bigr) \bigl( |\kvec| + \kvec \cdot \vvec_b - \ii \epsM \bigr) \bigl( |\kvec| + \kvec \cdot \vvec_b - \ii \epsP \bigr)} 
	\\ & \hspace{4cm}
	+ \frac{1 - \vvec_b \cdot \vvec_b}{\bigl( |\kvec| - \kvec \cdot \vvec_b - \ii \epsP \bigr) \bigl( |\kvec| + \kvec \cdot \vvec_b - \ii \epsM \bigr) \bigl( \ii \varepsilon \bigr)} 
	\biggr) 
	\per
    \nonumber
}
Here we have defined $\epsP = \epsF + \varepsilon$ and $\epsM = \epsF - \varepsilon$.  
Now it is clear that $\Ical_{ba}$ is identical to $\Ical_{ab}$ with the replacements $\vvec_a \to -\vvec_b$ and $\vvec_b \to - \vvec_a$; similarly $\Ical_{bb}$ is identical to $\Ical_{aa}$ with the replacements $\vvec_a \to \vvec_b$ and $\vvec_b \to \vvec_a$.  
It is straightforward to evaluate $\Ical_{aa}$ (or $\Ical_{bb}$) by working in spherical coordinates and taking the polar axis to align with $\vvec_a$ (or $\vvec_b$).  
However, since both $\vvec_a$ and $\vvec_b$ appear in the integrands of $\Ical_{ab}$ and $\Ical_{ba}$, an evaluation of these integrals for general velocities is more difficult.  
To simplify the problem, we now employ the assumptions in \eref{eq:model2_assume} and select the polar axis to align with $\vhat$.  
In spherical coordinates, the integrals become 
\bsa{}{
%---
	\Ical_{aa} & = \lim_{\epsF, \varepsilon \to 0^+} \frac{1}{2} \int_0^\infty \! \frac{k^2 \dd k}{2\pi^2} \frac{1}{\bigl( k - \ii \epsF \bigr)} \int_{-1}^{1} \frac{\dd \cos\theta}{2} \, \frac{1 - v_a^2}{\bigl( k + k v_a \cos \theta - \ii \epsP \bigr) \bigl( k + k v_a \cos \theta - \ii \epsM \bigr)} 
	\\ & \qquad 
	+ \lim_{\epsF, \varepsilon \to 0^+} \frac{1}{2} \int_0^\infty \! \frac{k^2 \dd k}{2\pi^2} \frac{1}{\bigl( \ii \varepsilon \bigr)} \int_{-1}^{1} \frac{\dd \cos\theta}{2} \, \frac{1 - v_a^2}{\bigl( k - k v_a \cos \theta - \ii \epsP \bigr) \bigl( k + k v_a \cos \theta - \ii \epsM \bigr)}  \,,
	\nonumber \\ 
%---
	\Ical_{ab} & = \lim_{\epsF, \varepsilon \to 0^+} \frac{1}{2} \int_0^\infty \! \frac{k^2 \dd k}{2\pi^2} \frac{1}{\bigl( k - \ii \epsF \bigr)} \int_{-1}^{1} \frac{\dd \cos\theta}{2} \, \frac{1 - v_a v_b}{\bigl( k + k v_a \cos \theta - \ii \epsP \bigr) \bigl( k + k v_b \cos \theta - \ii \epsP \bigr)} 
	\per
}
Evaluating the angular integral gives 
\bsa{}{
%---
	\Ical_{aa} & = \lim_{\epsF, \varepsilon \to 0^+} \frac{1}{2} \int_0^\infty \! \frac{k^2 \dd k}{2\pi^2} \frac{\bigl( 1 - v_a^2 \bigr) \log\frac{k (1+v_a) - \ii \epsP}{k (1-v_a) - \ii \epsP}}{2 k v_a \bigl( k - \ii \epsF \bigr) \bigl( \ii \varepsilon \bigr)} \\ 
%---
	\Ical_{ab} & = \lim_{\epsF, \varepsilon \to 0^+} \frac{1}{2} \int_0^\infty \! \frac{k^2 \dd k}{2\pi^2} \frac{\bigl( 1 - v_a v_b \bigr) \bigl( \log\frac{k (1+v_a) - \ii \epsP}{k (1-v_a) - \ii \epsP} - \log\frac{k (1+v_b) - \ii \epsP}{k (1-v_b) - \ii \epsP} \bigr)}{2 k \bigl( v_a - v_b \bigr) \bigl( k - \ii \epsF \bigr) \bigl( k - \ii \epsP \bigr)} 
	\per
}
Notice that the logs involving $\epsM$ have cancelled out.  
This is advantageous because $\epsP = \epsF + \varepsilon$ is always positive, but $\epsM = \epsF - \varepsilon$ could be either positive or negative, depending on how the limit $\epsF, \varepsilon \to 0^+$ is performed.  
The absence of $\epsM$ makes the limit unambiguous, and we are free to send $\epsF \to 0$ first, so that only $\varepsilon$ survives as the IR regulator.  
To extract the leading IR behavior, we can also expand in powers of $\varepsilon$ and regulate the integral by introducing an IR cutoff $k_\IR \equiv \varepsilon$ such that $k \geq k_\IR$.  
The $k$ integrals have UV divergences, which we regulate with a UV cutoff $k_\UV$ such that $k \leq k_\UV$.    
To leading order in powers of $k_\UV / k_\IR$ we find 
\bsa{}{
%---
	\Ical_{aa} 
    & = \lim_{k_\UV \to \infty} \lim_{k_\IR \to 0^+}
	\biggl( - \frac{\ii}{8 \pi^2} \biggr) \biggl( \frac{1 - v_a^2}{v_a} \biggr) \biggl( \log \frac{1 + v_a}{1 - v_a} \biggr) \biggl( \frac{k_\UV}{k_\IR} \biggr) 
    + \biggl( \frac{1}{4 \pi^2} \biggr) \biggl( \log \frac{k_\UV}{k_\IR} \biggr) \\ 
%---
	\Ical_{ab} 
    & = \lim_{k_\UV \to \infty} \lim_{k_\IR \to 0^+}
	\biggl( \frac{1}{8 \pi^2} \biggr) \biggl( \frac{1 - v_a v_b}{v_a - v_b} \biggr) \biggl( \log \frac{(1 + v_a) (1 - v_b)}{(1 - v_a) (1 + v_b)} \biggr) \biggl( \log \frac{k_\UV}{k_\IR} \biggr)
    \per
}
Note that $\Ical_{aa}$ and $\Ical_{bb}$ have linear divergences, whereas $\Ical_{ab}$ and $\Ical_{ba}$ have logarithmic divergences.  
Finally the vacuum persistence amplitude is given by \eref{eq:model2_W002_from_I} to be 
\bes{
    W_{0\to0}^{(2)} 
    = \lim_{k_\UV \to \infty} \lim_{k_\IR \to 0^+} \, (\qe)^2 \Biggl[
    & 
    \biggl( - \frac{\ii}{8 \pi^2} \biggr) \biggl( \frac{1 - v_a^2}{v_a} \biggr) \biggl( \log \frac{1 + v_a}{1 - v_a} \biggr) \biggl( \frac{k_\UV}{k_\IR} \biggr) 
    \\ & \ 
    - 2 \biggl( \frac{1}{8 \pi^2} \biggr) \biggl( \frac{1 - v_a v_b}{v_a - v_b} \log \frac{(1 + v_a) (1 - v_b)}{(1 - v_a) (1 + v_b)} - 2 \biggr) \biggl( \log \frac{k_\UV}{k_\IR} \biggr)
    \\ & \ 
    + \biggl( - \frac{\ii}{8 \pi^2} \biggr) \biggl( \frac{1 - v_b^2}{v_b} \biggr) \biggl( \log \frac{1 + v_b}{1 - v_b} \biggr) \biggl( \frac{k_\UV}{k_\IR} \biggr)
    \Biggr] 
    \per
}
The vacuum survival probability is calculated by $\Pbb_{0\to0} = \mathrm{exp}(\mathrm{Re} \, W_{0\to0}^{(2)})$ as in \eref{eq:P00_from_W002}.  
Since the terms with linear divergences are pure imaginary, they do not contribute to $\Pbb_{0\to0}$.  
The terms with a logarithmic UV divergence need to be renormalized, which will swap $k_\UV$ for a renormalization scale $\mu$, which is assumed to be $\mu > k_\IR$. 
The vacuum survival probability is found to be 
\bes{\label{eq:model2_P00}
    \Pbb_{0\to0} 
    & = \mathrm{exp}\Biggl[
    - \biggl( \frac{q^2 e^2}{2 \pi^2} \biggr) \biggl( \frac{1}{2} \frac{1 - v_a v_b}{v_a - v_b} \log \frac{(1 + v_a) (1 - v_b)}{(1 - v_a) (1 + v_b)} - 1 \biggr) \biggl( \log \frac{\mu}{k_\IR} \biggr)
    \Biggr] 
    \per
}
This represents the probability that a particle with charge $q$ that is abruptly decelerated from velocity $\vvec_a = v_a \, \vhat$ to velocity $\vvec_b = v_b \, \vhat$ \textit{does not} emit electromagnetic radiation.  
Note that the speed-dependent factor in parentheses is positive for all $0 \leq v_b < v_a < 1$, and so the probability is $0 < \Pbb_{0\to0} \leq 1$ as required. 
In the limit $v_b \to v_a$ the formula has $\Pbb_{0\to0} \to 1$, which agrees with the analysis in \sref{sec:model1}: a particle traveling at constant velocity does not radiate.  

%-------------------------------------------
% Single emission
%-------------------------------------------
\subsection{Single emission}
\label{sub:model2_emission}

%=========
An expression for the differential probability of a single photon emission was provided in \eref{eq:dP00g_from_jj}. 
Using the expression for $\tilde{j}(k)$ from \eref{eq:model2_current} leads to 
\ba{\label{eq:model2_dP00g}
	\dd\Pbb_{0\to0\gamma} 
	& = \lim_{\varepsilon \to 0^+} \bigl( - q^2 e^2 \Pbb_{0\to0} \bigr) 
    \frac{\dd^3 \pvec}{(2\pi)^3} \, \frac{1}{2E_p} \, \biggl( 
	\frac{\Ubb_a \cdot \Ubb_a}{\bigl( p \cdot \Ubb_a + \ii \varepsilon \bigr) \bigl( p \cdot \Ubb_a - \ii \varepsilon \bigr)} 
	- \frac{\Ubb_a \cdot \Ubb_b}{\bigl( p \cdot \Ubb_a + \ii \varepsilon \bigr) \bigl( p \cdot \Ubb_b + \ii \varepsilon \bigr)} 
	\nn & \hspace{3.5cm}
	- \frac{\Ubb_b \cdot \Ubb_a}{\bigl( p \cdot \Ubb_b - \ii \varepsilon \bigr) \bigl( p \cdot \Ubb_a - \ii \varepsilon \bigr)} 
	+ \frac{\Ubb_b \cdot \Ubb_b}{\bigl( p \cdot \Ubb_b - \ii \varepsilon \bigr) \bigl( p \cdot \Ubb_b + \ii \varepsilon \bigr)} \biggr) 
	\per
}
Recall that the four-vectors are $p^\mu = (E_p, \pvec)$, $\Ubb_a^\mu = (\gamma_a, \gamma_a \vvec_a)$, and $\Ubb_b^\mu = (\gamma_b, \gamma_b \vvec_b)$.  
If the parameter $\varepsilon$ takes a nonzero value, then the differential probability is free of poles throughout the full phase space $\pvec \in \Rbb^3$.  
If $\varepsilon = 0$ then there is a pole at $\pvec = 0$, corresponding to a soft IR divergence, and possibly also poles along the rays where $\pvec \parallel \vvec_{a,b}$ if $|\vvec_{a,b}| = 1$, corresponding to collinear IR divergences. 
For $0 \leq v_b \leq v_a < 1$ the collinear divergences are avoided, and we can set $\varepsilon = 0$ while remembering that it is necessary to reintroduce a regulator later to address the remaining soft IR divergence.  
Using $E_p = |\pvec|$ and $\phat = \pvec / |\pvec|$, the differential probability is expressed as 
\bes{\label{eq:model2_dP00g_3vec_v}
	\dd\Pbb_{0\to0\gamma} 
%---
	& = \biggl( \frac{q^2 e^2}{16 \pi^3} \Pbb_{0\to0} \biggr) 
    \biggl( \frac{\dd^3 \pvec}{|\pvec|^3} \biggr) 
    \biggl(
    \biggl| \frac{\vvec_a}{1 - \phat \cdot \vvec_a} - \frac{\vvec_b}{1 - \phat \cdot \vvec_b} \biggr|^2 
    - \frac{(\phat \cdot \vvec_a - \phat \cdot \vvec_b)^2}{(1 - \phat \cdot \vvec_a)^2 (1 - \phat \cdot \vvec_b)^2} 
    \biggr)
	\per
}
In the next subsection, we calculate the corresponding thermal pressure while exploring different choices for the parameters $\vvec_a$ and $\vvec_b$.  

%-------------------------------------------
% Implications for thermal pressure
%-------------------------------------------
\subsection{Implications for thermal pressure}
\label{sub:model2_pressure}

%=========
We now suppose that the charged particle's deceleration results from an increase in its mass across a planar domain wall.  
As its mass increases from $m_a$ to $m_b$, its velocity decreased from $\vvec_a$ to $\vvec_b$ accordingly.  
This situation is realized at a first order electroweak phase transition when quarks and charged leptons encounter a Higgs-phase bubble wall, leading to gluon and photon emission. 
We are interested in calculating the resultant thermal pressure $P_\mathrm{therm}$ acting on the wall.  
Since the thermal pressure is parametrically given by $P_\mathrm{therm} \sim \gamma_w T^3 \, \langle \Delta p_z \rangle$, as in \eref{eq:Ptherm_scaling}, the quantity of interest is the average longitudinal momentum transfer $\langle \Delta p_z \rangle$. 

%=========
To begin, let us first express the single photon emission probability in terms of masses and momenta.  
The three-momentum and energy of the incident particle are given by the relations $\pvec_a = \gamma_a m_a \vvec_a$ and $E_a = \gamma_a m_a$ where $\gamma_a = 1 / \sqrt{1 - |\vvec_a|^2}$ and $E_a = \sqrt{|\pvec_a|^2 + m_a^2}$.  
Similar expressions hold for the recoiling particle with the replacement of $a$ to $b$.  
Using this notation, the expression for $\dd\Pbb_{0\to0\gamma}$ from \eref{eq:model2_dP00g_3vec_v} is written equivalently as 
\bes{\label{eq:model2_dP00g_3vec_p}
	\dd\Pbb_{0\to0\gamma} 
%---
	& = \biggl( \frac{q^2 e^2}{16 \pi^3} \Pbb_{0\to0} \biggr) 
    \biggl( \frac{\dd^3 \pvec}{E_p} \biggr) 
    \Biggl( \ 
    \biggl| \frac{\pvec_a}{E_p E_a - \pvec \cdot \pvec_a} - \frac{\pvec_b}{E_p E_b - \pvec \cdot \pvec_b} \biggr|^2 
    \\ & \hspace{4cm}
    - \frac{(E_b \, \pvec \cdot \pvec_a - E_a \, \pvec \cdot \pvec_b)^2}{(E_p E_a - \pvec \cdot \pvec_a)^2 (E_p E_b - \pvec \cdot \pvec_b)^2} 
    \ \Biggr) 
	\com
}
where $E_p = |\pvec|$. 
Then the average longitudinal momentum transfer is calculated using \erefs{eq:avg_Delta_pz}{eq:Delta_pz}, which are reproduced here: 
\bes{\label{eq:avg_Delta_pz_sec6}
    \langle \Delta p_z \rangle = \int \! \dd \Pbb_{0 \to 0\gamma} \, \biggl( - \frac{|\pvec_{a,\perp}|^2 + m_a^2}{2 E_a} + \frac{|\pvec_{a,\perp} - \pvec_\perp|^2 + m_b^2}{2 (E_a - E_p)} + \frac{|\pvec_\perp|^2}{2 E_p} \biggr) 
    \per
}
Note that $\dd\Pbb_{0\to0\gamma}$ is a probability distribution over the photon momentum $\pvec$  where $\pvec_a$, $m_a$, $\pvec_b$, $m_b$, and $q$ are parameters.  
On the other hand, $\langle \Delta p_z \rangle$ is only a function of the parameters $\pvec_a$, $m_a$, and $m_b$ but not $\pvec_b$.  
This is because the derivation of $\Delta p_z$ in \eref{eq:Delta_pz} assumed on-shell energy-momentum relations as well as energy and transverse momentum conservation among the three particles, which fixed $\pvec_b$ in terms of $\pvec_a$ and $\pvec$.  
Conversely, the derivation of $\dd\Pbb_{0\to0\gamma}$ via the SCR formalism took the velocities $\vvec_a$ and $\vvec_b$ as arbitrary input parameters.  
In order to evaluate $\langle \Delta p_z \rangle$, we now must decide how to choose the various parameters: in particular how to select $\pvec_b$ when evaluating $\dd\Pbb_{0\to0\gamma}$.  
In the following subsections we explore three different choices.  

%-------------------------------------------
% choice \#1:  collinear motion
%-------------------------------------------
\subsubsection{choice \#1:  collinear motion}
\label{subsub:choice1}

%=========
As a first choice, suppose that the motion of particles $a$ and $b$ remains collinear, but otherwise the speeds are arbitrary.  
This means that $\vvec_a \parallel \vvec_b$ with $0 \leq v_b < v_a < 1$ as in \eref{eq:model2_assume}.  
With this restriction, the differential probability from \erefs{eq:model2_dP00g_3vec_v}{eq:model2_dP00g_3vec_p} leads to a simple expression, 
\bes{\label{eq:model2_dPdOmega}
	\frac{\dd\Pbb_{0\to0\gamma}}{\dd\Omega}
%---
	& = \frac{q^2 e^2}{16 \pi^3} \Pbb_{0\to0} \frac{\dd p}{p}  \frac{(\sin^2\theta) \, (v_a - v_b)^2}{(1 - v_a \, \cos\theta)^2 (1 - v_b \, \cos\theta)^2} 
	\com
}
which has been expressed in polar coordinates by using $\pvec = (\sin\theta \, \cos\phi \, , \, \sin\theta \, \sin\phi \, , \, \cos\theta) \, p$ and $\vvec_{a,b} = (0 \, , \, 0 \, , \, 1) \, v_{a,b}$ and $\dd^3 \pvec = p^2 \dd p \dd \Omega$.  
The emission vanishes along the axis of the charged particle ($\theta = 0$), but for $v_a \approx v_b \approx 1$ the emission strongly peaks in a forward cone where $\cos\theta \approx 1 - (1-v_a) (1-v_b) / (1 - v_a v_b)$. 

%=========
Here it is instructive to compare with the familiar results for bremsstrahlung radiation in classical electromagnetism.  
We can calculate the average energy output per unit angular frequency per unit solid angle as $E_p \dd\Pbb_{0\to0\gamma} / \dd \omega \dd \Omega$ where $E_p = \omega = p$.  
The corresponding quantity in classical electromagnetism can be calculated using Eq.~(14.67) of \rref{Jackson:1998nia} (note that $e_\mathrm{cgs}^2 = \alpha = e_\mathrm{HL}^2 / 4\pi$), and the two calculations yield equivalent results.  
Similarly, the average power received at a distant detector is given by Eq.~(14.39) of \rref{Jackson:1998nia} or Eq.~(11.74) of \rref{Griffiths:2024}, and an equivalent result can be derived from \eref{eq:model2_dP00g_3vec_v} if the duration of the acceleration cuts off the momentum integral.

\begin{figure}[t]
\centering \includegraphics[width=0.65\textwidth]{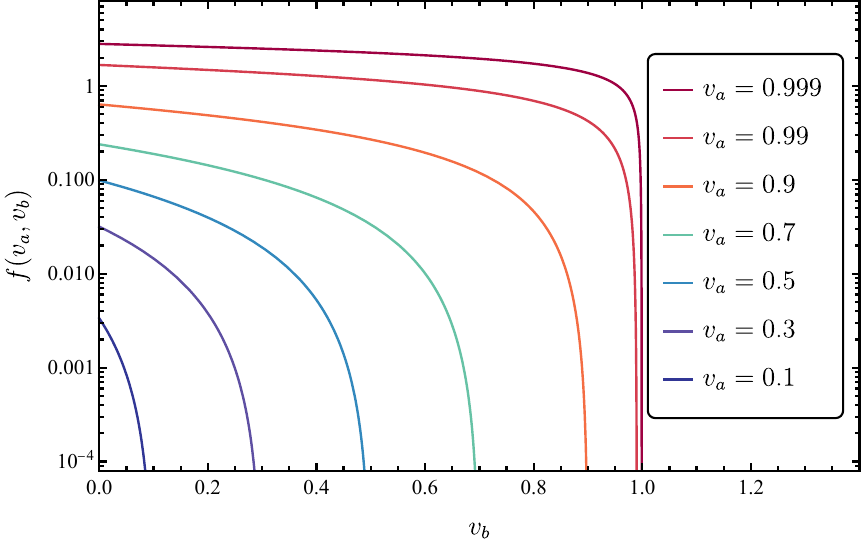}
\caption{\label{fig:model2_pdPdp} 
Speed dependence of the momentum spectrum.  In this calculation we assume that a charged particle decelerates abruptly from speed $v_a$ to speed $v_b < v_a$.  The momentum spectrum is found to be log-flat with an amplitude that depends on $v_a$ and $v_b$ as $p \, \dd \Pbb_{0\to0\gamma} / \dd p = ( q^2 e^2 \Pbb_{0\to0} / 2 \pi^2 ) \, f(v_a, v_b)$.  As $v_b$ approaches $v_a$ from below, the amplitude of the spectrum approaches zero as $(v_b - v_a)^2$, and there is no radiation.}
\end{figure}

%=========
Evaluating the angular integrals in \eref{eq:model2_dPdOmega} yields the single photon emission probability 
\bes{\label{eq:model2_pdPdp}
    p \frac{\dd \Pbb_{0\to0\gamma}}{\dd p} 
	= \biggl( \frac{q^2 e^2}{2 \pi^2} \Pbb_{0\to0} \biggr) 
	\, f(v_a, v_b) 
    \com
}
where we have defined the function 
\ba{\label{eq:model2_f_def}
	f(v_a, v_b) = 
	\frac{1}{2} \biggl( \frac{1 - v_a v_b}{v_a - v_b} \biggr) \log \frac{(1+v_a) (1-v_b)}{(1-v_a) (1+v_b)} - 1 
    \per 
}
A plot of $f(v_a, v_b)$ appears in \fref{fig:model2_pdPdp} and this quantity is positive for all $0 \leq v_b < v_a < 1$. We note that as $v_b\to v_a$, $f(v_a, v_b)\to 0$.

%=========
Since $p \dd \Pbb_{0\to0\gamma} / \dd p$ is independent of $p$, the momentum spectrum of photon radiation is log-flat. 
That is to say, there is equal probability for photon emission across logarithmically-spaced momentum intervals all the way from $\log p = - \infty$ to $\log p = + \infty$.  
If we calculate the total probability by integrating over momentum, there is a logarithmic IR divergence (toward $\log p = - \infty$) as well as a logarithmic UV divergence (toward $\log p = + \infty$).  
To handle these divergences we restrict the domain of integration to $p_\IR \leq p \leq p_\UV$ and then remove the regulators by sending $p_\IR \to 0$ and $p_\UV \to \infty$.  
Doing so gives the single photon emission probability: 
\bes{\label{eq:model2_P00g}
	\Pbb_{0\to0\gamma} 
%---
	& = \lim_{p_\IR \to 0} \lim_{p_\UV \to \infty}
	\biggl( \frac{q^2 e^2}{2 \pi^2} \Pbb_{0\to0} \biggr) 
	\, f(v_a, v_b) \, \log \frac{p_\UV}{p_\IR} 
    %+ O[(p_\UV/p_\IR)^0]
	\per
}
Although this expression is divergent as we remove the regulators, one can argue that we should instead set $p_\IR$ and $p_\UV$ equal to finite physical scales.  
In practice, very soft emission with energy below the threshold of the experiment ($p < \Delta p_\mathrm{det}$) cannot be detected; see the discussion in \aref{app:cancellation}.  
This motivates selecting the IR cutoff to be the experimental energy resolution, $p_\IR = \Delta p_\mathrm{det}$.  
For electromagnetic radiation in a medium, the plasma frequency may also enter to cutoff the would-be IR divergence.  
As for the UV cutoff, the whole calculation is predicated on assuming that the radiation does not significantly back react on the emission of the current from which it is sourced.\footnote{The same treatment is often employed in studies of classical electromagnetism.  The source charge and current densities are assumed to be fixed externally, and Maxwell's equations are solved to calculate the resultant radiation.}  
However, if the radiation carries away an energy that is larger than the energy available in the radiating particle, then the assumption of negligible back reaction is invalid.  
This argument motivates selecting the UV cutoff to be the difference of the incident and recoiling particles' energies $p_\UV = E_a - E_b$.  
In \sref{sec:model3} we will see that the duration of the deceleration introduces another UV cutoff, which can be smaller than the energy difference.  

%=========

%=========
We evaluate the average longitudinal momentum transfer by using the formula for $\langle \Delta p_z \rangle$ from \eref{eq:avg_Delta_pz_sec6}.  
Now it is convenient to take $\pvec_{a,\perp} = {\bm 0}$.  
The angular integrals are free of divergences, but the integral over $p$ has a linear UV divergence, which we regulate by imposing $p \leq p_\UV$ and sending $p_\UV \to \infty$. 
We also write $\vvec_a = \pvec_a / E_a$ and expand to leading order in large $E_a$ while holding $v_b$ fixed.  
After evaluating the integrals, we are left with 
\bes{\label{eq:model2_choice1_Dpz}
    \langle \Delta p_z \rangle 
    & \approx \biggl( \frac{q^2 e^2}{4 \pi^2} \Pbb_{0\to0} \biggr) \biggl( \frac{1 - v_b}{v_b^2} \biggr) \biggl( 1 - \frac{1 - v_b^2}{2 v_b} \log \frac{1 + v_b}{1 - v_b} \biggr) \, p_\UV + O(E_a^{-1})
    \per
}
Note that the average longitudinal momentum transfer is found to be $O(E_a^0)$ at the leading order, but it also scales as $p_\UV^1$.  
In the preceding paragraph, we discussed how the energy difference of particles $a$ and $b$ is a well-motivated choice for the UV cutoff: $p_\UV \sim E_a - E_b$.  
If $\pvec_b$ is held fixed as we send $E_a \to \infty$ then $p_\UV \sim E_a^1$.  
To track the scaling with the Lorentz factor $\gamma_w$ we can write $E_a \sim \gamma_w T$.  
This implies $\langle \Delta p_z \rangle \propto \gamma_w^1$ and therefore $P_\mathrm{therm} \propto \gamma_w^2$.  
On the other hand, if a different scale enters as the UV cutoff, which is independent of $\gamma_w$, then we would find instead $\langle \Delta p_z \rangle \propto \gamma_w^0$ and $P_\mathrm{therm} \propto \gamma_w^1$.  
We revisit this point when we discuss a model with finite bubble wall thickness in \sref{sec:model3}.  

%=========
From the expression for $\langle \Delta p_z \rangle$ in \eref{eq:model2_choice1_Dpz}, we see that the average longitudinal momentum transfer vanishes as $v_b$ approaches $v_a = 1$ from below.  
This observation suggests that we should take care in selecting $\pvec_b$, particularly if we are interested in the $E_a \to \infty$ limit.  
For example, if we choose $\pvec_b$ such that it approaches $\pvec_a$ as $E_a \to \infty$, then we expect to find an additional suppression of the longitudinal momentum transfer.  
Next we consider a choice for $\pvec_b$ with precisely this behavior. 

%-------------------------------------------
% choice \#2:  impose 1-to-1 kinematics
%-------------------------------------------
\subsubsection{choice \#2:  impose 1-to-1 kinematics}
\label{subsub:choice2}

%=========
As a second choice, we suppose that $\pvec_b$ is related to $\pvec_a$ according to the kinematic relations that one expects for a 1-to-1 transition.  
In the regime where the radiation does not back react significantly on the motion of the charged radiating particles, then it is straightforward to calculate $\pvec_b$ in terms of $\pvec_a$ assuming that the momentum change results from a mass increase, $m_a$ to $m_b$, at a planar bubble wall~\cite{Bodeker:2009qy}.  
In the rest frame of the wall, energy and transverse momentum are conserved (since the background is static and transverse-homogeneous), but longitudinal momentum is not conserved (since the wall breaks longitudinal translation invariance).  
It follows that the energy and transverse momentum are unchanged, while the longitudinal momentum decreases~\cite{Bodeker:2009qy}:
\bes{\label{eq:choice2}
    E_b = E_a 
    \ , \quad 
    \pvec_{b,\perp} = \pvec_{a,\perp} 
    \ , \quad \text{and} \quad 
    p_{b,z} = \sqrt{ E_b^2 - |\pvec_{b,\perp}|^2 - m_b^2 } \approx p_{a,z} - \frac{m_b^2 - m_a^2}{2 E_a} + O(E_a^{-2}) 
    \per 
}
In the approximate expression for $p_{b,z}$, we assume that the longitudinal momentum of the incident particle is large compared to the masses and transverse momentum: $p_{a,z} \approx E_a \gg m_a, m_b, |\pvec_{a,\perp}|$. 
Note that sending $m_b \to m_a$ implies $\pvec_b \to \pvec_a$ and therefore also $v_b \to v_a$ and $\dd\Pbb_{0\to0\gamma} \to 0$.

%=========
The single photon emission probability can be calculated from either \eref{eq:model2_dP00g_3vec_p} or \eref{eq:model2_pdPdp} upon making the substitutions in \eref{eq:choice2}.  
To leading order in large $E_a$, we find 
\bes{\label{eq:model2_pdPdp_choice2}
    p \frac{\dd \Pbb_{0\to0\gamma}}{\dd p} 
	= \biggl( \frac{q^2 e^2}{2 \pi^2} \Pbb_{0\to0} \biggr) 
	\, \biggl( \frac{m_b^2 + m_a^2}{m_b^2 - m_a^2} \, \log \frac{m_b}{m_a} - 1 \biggr) + O(E_a^{-2})
    \per
}
Several remarks are in order.  
First, one arrives at the same result by following the quantum particle splitting formalism developed in BM17. 
For comparison, we present this alternative calculation in \aref{app:comparison}.  
Note that the derivation in BM17 and \aref{app:comparison} assumes that the emitted photon is soft (since it uses the splitting function for soft vector emission), and therefore we expect to find agreement with the SCR formalism when $\pvec_b$ is chosen according to \eref{eq:choice2}.  
This agreement can be taken as evidence that the SCR formalism can provide a reliable description of the system, although care must be taken in selecting $\pvec_b$.  
Second, if the masses are hierarchical $0 < m_a \ll m_b$ then the spectrum is enhanced by a large logarithm. 
On the other hand, if the mass does not change significantly then 
\ba{
	\lim_{m_b \to m_a} p \frac{\dd \Pbb_{0\to0\gamma}}{\dd p} \sim \biggl( \frac{q^2 e^2}{2 \pi^2} \Pbb_{0\to0} \biggr) \biggl( \frac{1}{3} \frac{(m_b - m_a)^2}{m_a^2} \biggr) + O(E_a^{-2})
	\com
}
and the spectrum is suppressed by the tiny mass difference. 
Third, the spectrum is found to be log-flat in the momentum of the emitted photon.  
This is the same behavior that we encountered in \eref{eq:model2_pdPdp}, which should be no surprise, since all that has changed between \eref{eq:model2_pdPdp} and \eref{eq:model2_pdPdp_choice2} is the restriction that $v_b$ should be linked to $v_a$ through \eref{eq:choice2}. 

%=========
We evaluate the average longitudinal momentum transfer by using the formula for $\langle \Delta p_z \rangle$ from \eref{eq:avg_Delta_pz_sec6}.  
Now it is convenient to take $\pvec_{a,\perp} = {\bm 0}$.  
Note that the integrals over the angular coordinates are well behaved, but the integral over the photon momentum has logarithmic IR and UV divergences.  
To handle these divergences we restrict the domain of integration to $p_\IR \leq p \leq p_\UV$.  
Doing so gives 
\bes{\label{eq:model2_choice2_Dpz}
    \langle \Delta p_z \rangle 
    & = \biggl( \frac{q^2 e^2}{4 \pi^2} \Pbb_{0\to0} \biggr) \biggl( \log \frac{m_b}{m_a} - \frac{m_b^2 - m_a^2}{m_b^2 + m_a^2} \biggr) \biggl( \frac{m_b^2 + m_a^2}{E_a} \biggr) \log \frac{p_\UV}{p_\IR} + O(E_a^{-2})
    \com
}
where we only show the leading behavior at large $E_a$.  
Note that $\langle \Delta p_z \rangle \propto E_a^{-1}$, and this behavior should be contrasted with \eref{eq:model2_choice1_Dpz} that went as $E_a^{0} \times p_\UV$.  
Here as we raise $E_a$ we also send $\pvec_b \to \pvec_a$ according to \eref{eq:choice2}, which leads to a suppression.  
To understand the dependence on the Lorentz factor $\gamma_w$, we take $E_a \sim \gamma_w T$, which gives $\langle \Delta p_z \rangle \propto \gamma_w^{-1}$ and therefore the thermal pressure scales as $P_\mathrm{therm} \propto \gamma_w^0$.  
This result is in line with a remark at the end of sec.~4 of BM17, where the authors note that the channel with a massive radiator producing massless radiation will contribute to the thermal pressure as $\langle \Delta p_z \rangle \propto \gamma_w^{-1}$ and $P_\mathrm{therm} \propto \gamma_w^{0}$. 

%-------------------------------------------
% choice \#3:  impose 1-to-2 kinematics
%-------------------------------------------
\subsubsection{choice \#3:  impose 1-to-2 kinematics}
\label{subsub:choice3}

%=========
As a third choice, we adopt the following procedure to replace $\pvec_b$ with a function of $\pvec$.  
It is worth emphasizing that $\dd\Pbb_{0\to0\gamma}$ is a probability distribution over $\pvec$ where $\pvec_a$ and $\pvec_b$ are parameters of the distribution.  
Allowing $\pvec_b$ to depend upon $\pvec$ mixes the distribution's parameters and independent variables. 
We explore this approach in order to compare with the analysis of HKLTW20, and because it is motivated by seeking to account for the backreaction of the radiation on the radiating particles while ensuring energy and transverse momentum are conserved between the three particles.

%=========
We replace $\pvec_b$ with a function of $\pvec$ according to 
\bes{\label{eq:choice3}
    E_b = E_a - E_p 
    \ , \qquad 
    \pvec_{b,\perp} = \pvec_{a,\perp} - \pvec_\perp 
    \ , \qquad \text{and} \qquad 
    p_{b,z} = \sqrt{ E_b^2 - |\pvec_{b,\perp}|^2 - m_b^2 } 
    \per 
}
These relations enforce energy and transverse momentum conservation between the incident particle (momentum $\pvec_a$), the recoiling particle (momentum $\pvec_b$), and the radiation (momentum $\pvec$).  
We also enforce an on-shell energy-momentum relation for all three particles.  
With only those restrictions, the longitudinal momentum need not be conserved ($p_{b,z} \neq p_{a,z} - p_z$).  
The absence of longitudinal momentum conservation is expected for a particle that passes through a bubble where its mass changes from $m_a$ to $m_b$.  
As we remove the effect of the wall by sending $m_b \to m_a$, we would like to see that \eref{eq:choice3} only has real solutions for $\pvec = 0$ such that it additionally enforces $p_{b,z} = p_{a,z}$.  
Then there would be no radiation since the single photon emission probability vanishes if $m_b = m_a$ and $\pvec_b = \pvec_a$; see \eref{eq:model2_dP00g_3vec_p}.  
However, setting $m_b = m_a$ in \eref{eq:choice3} does not imply $\pvec_b = \pvec_a$; instead, there are real solutions that do not exhibit longitudinal momentum conservation.   
Therefore the probability does not vanish in the limit $m_b \to m_a$, as one should expect, since it still allows $\vvec_b \neq \vvec_a$.  
This behavior was first discussed by the authors of \rref{Azatov:2020ufh} and later recognized in v3 of HKLTW20. 

%=========
The single photon emission probability can be calculated from \eref{eq:model2_dP00g_3vec_p} upon making the substitutions in \eref{eq:choice3}.  
To leading order in large $E_a$, the differential probability is 
\bes{
    p \frac{\dd \Pbb_{0\to0\gamma}}{\dd p \, \dd \Omega} \approx \biggl( \frac{q^2 e^2}{4 \pi^3} \Pbb_{0\to0} \biggr) \frac{1}{\sin^2 \theta} + O(E_a^{-1})
    \per
}
As anticipated by the discussion above, the spectrum does not vanish in the limit $m_b \to m_a$.  
The spectrum is once again found to be log-flat in the photon momentum $p$.  
Additionally, there is a singularity at $\theta = 0,\pi$, although this behavior is an artifact of the large $E_a$ expansion.  
If the large $E_a$ expansion has not been performed then the angular integrals would lead to a logarithm $\sim \log m_b / m_a$.  

%=========
We evaluate the average longitudinal momentum transfer by using the formula for $\langle \Delta p_z \rangle$ from \eref{eq:avg_Delta_pz_sec6}.  
Upon weighting the probability by $\Delta p_z$, the angular integrals are well behaved, and the momentum integral has a linear UV divergence.  
We regulate the divergence by resticting the domain of integration to $0 \leq p \leq p_\UV$.  
To leading order in large $E_a$, the average longitudinal momentum transfer is found to be 
\bes{\label{eq:model2_choice3_avg_Delta_pz}
    \langle \Delta p_z \rangle 
    & \approx \biggl( \frac{q^2 e^2}{2 \pi^2} \Pbb_{0\to0} \biggr) \, p_\UV + O(E_a^{-1})
    \per
}
If we write $p_\UV \sim E_a \sim \gamma_w T$ and consider the limit $\gamma_w \to \infty$, this calculation reveals that the average longitudinal momentum transfer is enhanced, diverging as $\langle \Delta p_z \rangle \propto \gamma_w^{1}$.  
This result is in line with the analytical and numerical results presented in HKLTW20.  
For example, compare eq.~(4.39) or table~2 of HKLTW20 with \eref{eq:model2_choice3_avg_Delta_pz} and note that $e^2/2\pi^2 = 2 \alpha / \pi \approx 0.0047$. 
If the momentum transfer scales as $\langle \Delta p_z \rangle \propto \gamma_w^{1}$, then the thermal pressure scales as $P_\mathrm{therm} \propto \gamma_w^{2}$, in agreement with HKLTW20.  

%-------------------------------------------
% Limitations of abrupt deceleration
%-------------------------------------------
\subsection{Limitations of abrupt deceleration}
\label{sub:model2_limitation}
%=========
The preceding estimates assume that the incident particle energy $\sim \gamma_w T$ acts as the physical UV cutoff.  
However, this assumption of an arbitrarily abrupt deceleration may not be reliable for calculating the very high-momentum end of the spectrum.  
One should expect the high-momentum photons to be sensitive to the scale and shape of the bubble wall, \textit{i.e.} the duration of the deceleration.\footnote{Some intuition is gained by considering nuclear form factors, which suppress scatterings with momentum transfer larger than the inverse nucleon length scale $|\qvec| > 1/R_n$.  For example see Ch.~7 of \rref{Thomson:2013zua}.  }  
This discussion suggests that the wall-crossing time $\approx L_w / c$ (in the wall's rest frame) will enter as a UV cutoff, and if $1/L_w < \gamma_w T$ then the thermal pressure would instead scale like $P_\mathrm{therm} \propto \gamma_w T^3 / L_w$.  
We explore this idea with a concrete calculation in \sref{sec:model3}.  

%==================================
% Model 3:~~Gradual deceleration
%==================================
\section{Model 3:~~Gradual deceleration}
\label{sec:model3}

%=========
As a final scenario, we consider a gradual deceleration from velocity $\vvec_a$ to velocity $\vvec_b$.  
We expect that softening the abrupt deceleration from \sref{sec:model2} will modify the predicted spectrum of radiation at high photon momentum. 
A similar conclusion has been drawn by other authors in previous studies~\cite{BarrosoMancha:2020fay,Azatov:2020ufh,Gouttenoire:2021kjv}, who find that the nonzero wall thickness $L_w$ introduces a UV cutoff $p_\UV \propto L_w^{-1}$. 

%-------------------------------------------
% Current density
%-------------------------------------------
\subsection{Current density}
\label{sub:model3_current}

%=========
Suppose that the charged particle initially travels with velocity $\vvec_a$, that its velocity temporarily decreases linearly with proper time, and that it finally travels with velocity $\vvec_b$.  
The four-velocity of the charged particle can be written as 
\ba{
	\Ubb^\mu(\tau) = \begin{cases}
	\Ubb_a^\mu = \bigl( \gamma_a \, , \, \gamma_a \vvec_a \bigr) & , \quad \tau < \tau_0 -\tfrac{\tau_w}{2} \\ 
	\Ubb_w^\mu(\tau) = \bigl( \gamma_w(\tau) \, , \, \gamma_w(\tau) \vvec_w(\tau) \bigr) & , \quad \tau_0 - \tfrac{\tau_w}{2} < \tau < \tau_0 + \tfrac{\tau_w}{2} \\ 
	\Ubb_b^\mu = \bigl( \gamma_b \, , \, \gamma_b \vvec_b \bigr) & , \quad \tau_0 + \tfrac{\tau_w}{2} < \tau 
	\end{cases}
	\com
}
where the velocity of the particle as it transitions through the wall is
\ba{
	\vvec_w(\tau) = \vvec_a - \bigl( \vvec_a - \vvec_b \bigr) \frac{\tau - \tau_0 + \tau_w/2}{\tau_w} 
	\per
}
This behavior is intended to model the dynamics of a particle that is incident upon a bubble wall where its mass increases  and its speed decreases.  
Rather than modelling how the mass varies across the wall, we simplify the problem by assuming that the deceleration is linear in proper time.  
We expect that our qualitative conclusions regarding how radiation depends on the wall thickness (here parametrized by $\tau_w$) are insensitive to the shape of the interpolating function.  
The particle reaches the wall at $\tau = \tau_0 - \tau_w / 2$, it crosses the middle of the wall at $\tau = \tau_0$, and it has fully passed through the wall at $\tau = \tau_0 + \tau_w / 2$.  
The Lorentz factors are $\gamma_{a,b,w} = 1 / \sqrt{1 - |\vvec_{a,b,w}|^2}$, which ensure that $\Ubb_{a,b,w} \cdot \Ubb_{a,b,w} = 1$.  
In other words, the radiating particle remains on-shell (with respect to the varying local mass) as it crosses the wall.  
As in \sref{sec:model2} we assume that the motion remains collinear; see \eref{eq:model2_assume}.  
The corresponding worldline is given by integrating the four-velocity
\ba{
	\Xbb^\mu(\tau) = \Xbb_0^\mu + \int_{\tau_0}^\tau \! \dd \tau^\prime \, \Ubb^\mu(\tau^\prime) 
	\com
}
where we impose the initial condition $\Xbb^\mu(\tau) = \Xbb_0^\mu$ at $\tau = \tau_0$.  
We can express the worldline as 
\ba{\label{eq:model3_Xmu}
	\Xbb^\mu(\tau) = \begin{cases}
	\Xbb_a^\mu(\tau) \equiv \bigl( T_a(\tau) \, , \, \Xvec_a(\tau) \bigr) & , \quad \tau < \tau_0 - \tfrac{\tau_w}{2} \\ 
	\Xbb_w^\mu(\tau) \equiv \bigl( T_w(\tau) \, , \, \Xvec_w(\tau) \bigr) & , \quad \tau_0 - \tfrac{\tau_w}{2} < \tau < \tau_0 + \tfrac{\tau_w}{2} \\ 
	\Xbb_b^\mu(\tau) \equiv \bigl( T_b(\tau) \, , \, \Xvec_b(\tau) \bigr) & , \quad \tau_0 + \tfrac{\tau_w}{2} < \tau 
	\end{cases}
	\com
}
where the expression for $T_w(\tau)$ is very lengthy, and the other expressions are  
\bsa{}{
	\Xvec_w(\tau) & =  \Xvec_0 + \tfrac{\tau_w}{v_a - v_b} \biggl[ \sqrt{1 - \Bigl( v_a - \tfrac{(v_a - v_b) (\tau - \tau_0 + \tau_w/2)}{\tau_w} \Bigr)^2} - \sqrt{1 - \Bigl( \tfrac{v_a + v_b}{2} \Bigr)^2} \biggr] \vhat \\ 
	\Xbb_a^\mu(\tau) & = \Xbb_w^\mu(\tau_0 - \tfrac{\tau_w}{2}) + \bigl( \tau - \tau_0 + \tfrac{\tau_w}{2} \bigr) \, \Ubb_a^\mu \\ 
	\Xbb_b^\mu(\tau) & = \Xbb_w^\mu(\tau_0 + \tfrac{\tau_w}{2}) + \bigl( \tau - \tau_0 - \tfrac{\tau_w}{2} \bigr) \, \Ubb_b^\mu 
	\per
}
Note that $\Xbb^\mu(\tau)$ is continuous at $\tau = \tau_0 - \tau_w/2$ and at $\tau_0 + \tau_w/2$.  
The wall thickness $L_w$ is given by 
\ba{
	L_w = \Bigl( \Xvec_w(\tau_0 + \tau_w/2) - \Xvec_w(\tau_0 - \tau_w/2) \Bigr) \cdot \vhat = \frac{\sqrt{1-v_b^2} - \sqrt{1 - v_a^2}}{v_a - v_b} \, \tau_w 
	\com
}
which allows us to exchange $\tau_w$ and $L_w$ in the formulas.  
If the two speeds are close, then $L_w = \gamma_a v_a \tau_w + O(v_b - v_a)$.  
The Fourier transform of the current density is calculated by evaluating the integral in \eref{eq:jmuFT}.  
Doing so gives 
\bsa{eq:model3_current}{
	\tilde{j}^\mu(k) & = \tilde{j}_a^\mu(k) + \tilde{j}_w^\mu(k) + \tilde{j}_b^\mu(k) \\ 
	\tilde{j}_a^\mu(k) & = \lim_{\varepsilon \to 0^+} (- \ii) \qe \, \ee^{\ii k \cdot \Xbb_w(\tau_0-\tau_w/2)} \, \frac{\Ubb_a^\mu}{k \cdot \Ubb_a - \ii \varepsilon} \\ 
	\tilde{j}_w^\mu(k) & = \qe \int_{\tau_0-\tau_w/2}^{\tau_0+\tau_w/2} \! \dd\tau \, \Ubb_w^\mu(\tau) \, \ee^{\ii k \cdot \Xbb_w(\tau)} \\ 
	\tilde{j}_b^\mu(k) & = \lim_{\varepsilon \to 0^+} (+ \ii) \qe \, \ee^{\ii k \cdot \Xbb_w(\tau_0+\tau_w/2)} \, \frac{\Ubb_b^\mu}{k \cdot \Ubb_b + \ii \varepsilon} 
	\per
}

%-------------------------------------------
% Single emission
%-------------------------------------------
\subsection{Single emission}
\label{sub:model3_emission}

%=========
An expression for the differential probability of a single photon emission was provided in \eref{eq:dP00g_from_jj}. 
Using the expression for $\tilde{j}(k)$ from \eref{eq:model3_current} leads to
\bes{\label{eq:model3_dP}
%---
	\dd\Pbb_{0\to0\gamma} 
	& = \dd\Pbb_{0\to0\gamma}^{(a,b)} + \dd\Pbb_{0\to0\gamma}^{(w)} \\ 
%---
	\dd\Pbb_{0\to0\gamma}^{(a,b)} 
	& = - \Pbb_{0\to0} \frac{\dd^3 \pvec_c}{(2\pi)^3} \, \frac{1}{2E_c} \, \Bigl( 
	\tilde{j}_a \cdot \tilde{j}_a^\ast 
	+ \tilde{j}_a \cdot \tilde{j}_b^\ast  
	+ \tilde{j}_b \cdot \tilde{j}_a^\ast 
	+ \tilde{j}_b \cdot \tilde{j}_b^\ast 
	\Bigr) \\ 
%---
	\dd\Pbb_{0\to0\gamma}^{(w)} 
	& = - \Pbb_{0\to0} \frac{\dd^3 \pvec_c}{(2\pi)^3} \, \frac{1}{2E_c} \, \Bigl( 
	\tilde{j}_a^\ast \cdot \tilde{j}_w 
	+ \tilde{j}_b^\ast \cdot \tilde{j}_w 
	+ \tilde{j}_w^\ast \cdot \tilde{j}_a 
	+ \tilde{j}_w^\ast \cdot \tilde{j}_b 
	+ \tilde{j}_w^\ast \cdot \tilde{j}_w 
	\Bigr) 
	\per
}
The calculation of $\dd\Pbb_{0\to0\gamma}^{(a,b)}$ runs parallel to the calculation of $\dd\Pbb_{0\to0\gamma}$ in \sref{sec:model2}.  
The key difference is the factors of $\ee^{\ii k \cdot \Xbb_w(\tau_0-\tau_w/2)}$ and $\ee^{\ii k \cdot \Xbb_w(\tau_0+\tau_w/2)}$ that appear in \eref{eq:model3_current}.  
These phase factors will cancel on the ``diagonal'' terms $\tilde{j}_a^\ast \cdot \tilde{j}_a$ and $\tilde{j}_b^\ast \cdot \tilde{j}_b$, but they will  contribute on the ``cross'' terms $\tilde{j}_a^\ast \cdot \tilde{j}_b$ and $\tilde{j}_b^\ast \cdot \tilde{j}_a$.  
So we can simply use the result from \sref{sub:model2_emission} and add these extra factors.  
After performing the angular integrals and removing the IR regulator ($\varepsilon \to 0$) the momentum spectrum of radiation is found to be 
\bes{\label{eq:model3_pdPdp}
	& p \frac{\dd\Pbb_{0\to0\gamma}}{\dd p} 
	= \biggl( \frac{q^2 e^2}{2 \pi^2} \Pbb_{0\to0} \biggr) \, f(v_a, v_b, p) + p \frac{\dd\Pbb_{0\to0\gamma}^{(w)}}{\dd p} 
	\com
}
where 
\ba{
	& 1 + f(v_a, v_b, p) = 
	\frac{1}{4} \biggl( \frac{1 - v_a v_b}{v_a - v_b} \biggr) 
	\biggl[ 
	\mathrm{Ei}\Bigl( \tfrac{- \ii L_w p (1+v_a)}{v_a} \Bigr) \ee^{\ii p L_w / v_a} 
	- \mathrm{Ei}\Bigl( \tfrac{- \ii L_w p (1-v_a)}{v_a} \Bigr) \ee^{\ii p L_w / v_a} 
	\\ & \quad 
	+ \mathrm{Ei}\Bigl( \tfrac{- \ii L_w p (1-v_b)}{v_b} \Bigr) \ee^{\ii p L_w / v_b} 
	- \mathrm{Ei}\Bigl( \tfrac{- \ii L_w p (1+v_b)}{v_b} \Bigr) \ee^{\ii p L_w / v_b} 
	\biggr] \ee^{-\ii p [ T_w(\tau_0+\tau_w/2) - T_w(\tau_0-\tau_w/2) ]} + \cc 
	\com
	\nonumber 
}
and $\mathrm{Ei}(z) = - \int_{-z}^\infty \dd t \, \ee^{-t}/t$ is the exponential integral function.  
Notice that the momentum spectrum is no longer log-flat, as we saw in \sref{sec:model2}, but rather the factor $f$ carries an additional momentum dependence.  
It is illuminating to consider the limit of small $L_w p$ in which 
\ba{
	f(v_a, v_b, p) = f(v_a, v_b) \Bigl[ 1 - \alpha(v_a,v_b) \, \bigl( L_w p \bigr)^2 + O[(L_w p)^3] \Bigr] 
    \qquad \text{as $L_w p \to 0$}
	\com
}
where $f(v_a, v_b)$ was defined in \eref{eq:model2_pdPdp} and where $\alpha(v_a,v_b)$ is a complicated function of the two speeds.  
This means that the spectrum is not flat in $\log p$ for all momenta, but rather it is only log-flat up to $p \approx 1 / \sqrt{\alpha(v_a,v_b)} L_w$.  
In other words, the wall thickness enters as a UV cutoff, as expected, and there is an additional velocity dependence through $\alpha(v_a,v_b)$.  
In this sense, we can interpret the ratio $f(v_a,v_b,p) / f(v_a,v_b)$ as the `form factor,' which accounts for the suppression of high-momentum radiation due to the finite wall thickness.  
For large values of $L_w p$, we find $f(v_a, v_b, p) < 0$ corresponding to a negative $\Pbb_{0\to0\gamma}^{(a,b)}$.  
In this regime, the contribution from $\Pbb_{0\to0\gamma}^{(w)}$ can no longer be neglected, and it must enter with a comparable magnitude to ensure that $\Pbb_{0\to0\gamma} = \Pbb_{0\to0\gamma}^{(a,b)} + \Pbb_{0\to0\gamma}^{(w)}$ is positive.  
We investigate this regime numerically in the following subsection.  

%-------------------------------------------
% Numerical validation
%-------------------------------------------
\subsection{Numerical validation}
\label{sub:model3_numerical}

%=========
To explore the full parameter space, we use numerical methods to evaluate the differential probability $p \, \dd \Pbb_{0\to0\gamma} / \dd p / \dd \cos\theta$ for single photon emission directly from \eref{eq:model3_dP}.  
Here $p$ is the photon's momentum and $\theta$ is the angle that it makes with the trajectory of the charged particle.  

%=========
\begin{figure}[t]
\centering 
\includegraphics[width=0.65\textwidth]{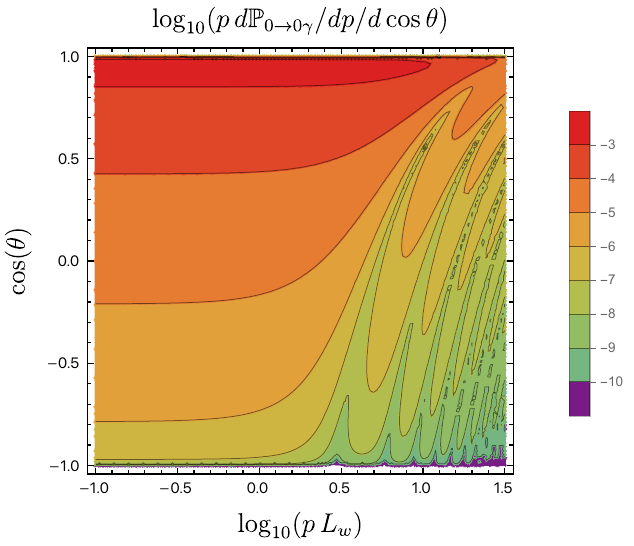} 
\caption{Distribution of photon radiation for the gradual deceleration model.  A charged particle with velocity $\vvec_a = v_a \, \vhat$ decelerates to velocity $\vvec_b = v_b \, \vhat$ over a distance $L_w$, and this figure shows the probability distribution to emit a single photon with momentum $p$ at an angle $\theta$ from the path of the particle.  In this example we have taken $v_a = 0.9$ and $v_b = 0.8$, and the qualitative behavior is insensitive to these parameters.}
\label{fig:model3_fig3}
\end{figure}

%=========
\begin{figure}[t]
\centering 
\includegraphics[width=\textwidth]{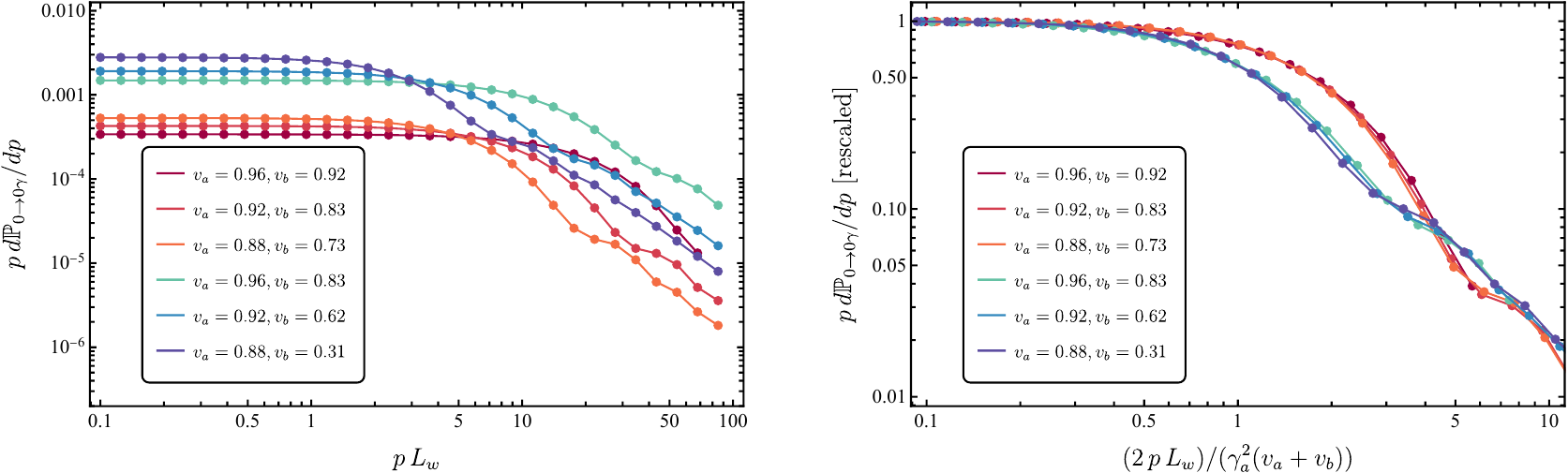} 
\caption{Momentum spectrum of photon radiation for the gradual deceleration model.  The curves correspond to different choices for the incident and recoiling particle speeds, $v_a$ and $v_b$ respectively. The left panel shows the differential probability to emit a photon with momentum between $\ln p$ and $\ln p + \dd \ln p$ where $p$ is expressed in units of the wall thickness $L_w$.  In the right panel, the horizontal axis has been scaled by a factor that depends on the speeds to illustrate that emission with $p \gtrsim \gamma_a^2 (v_a + v_b) / 2 L_w$ is suppressed. The vertical axis of the right plot has been normalized to one.
}
\label{fig:model3_fig4}
\end{figure}

%=========
The plot in \fref{fig:model3_fig3} illustrates the full dependence on $p$ and $\cos\theta$.  
The contours are surfaces of constant $p \, \dd \Pbb_{0\to0\gamma} / \dd p / \dd \cos\theta$ with logarithmic spacing in factors of $10$.  
The largest values are found in the upper-left corner (red), and the smallest values in the lower-right corner (green).  
Since the function is largest near to $\cos\theta \approx 1$, it implies that the emission is strongly peaked on a forward cone around the radiating particle.  
This behavior is familiar from bremstrahlung radiation in classical electromagnetism, which has maximal emission at $\cos\theta \approx 1 - (1-v)/4$ \cite{Jackson:1998nia}.  
As a function of momentum, the spectrum is insensitive to $p$ for $p \lesssim 1 / L_w$.  
This behavior is compatible with the log-flat momentum distribution that we have seen previously.  
Toward larger momentum, $p \gtrsim 1 / L_w$, the spectrum tends to decrease, but it also oscillates, as indicated by the banded features in the lower-right corner.  
The decreasing spectrum indicates that $1/L_w$ enters as a UV cutoff, and the emission is suppressed for larger momentum; we discuss this point further below.  
As for the oscillations, in general, the Fourier transform of a function with a kink or discontinuity exhibits oscillations toward large Fourier variable (\textit{i.e.}, frequency, wavenumber).  
Presumably, the oscillatory momentum spectrum is linked to the kinks in $\Xbb^\mu(\tau)$ our model in \eref{eq:model3_Xmu}.  
If we had modeled the gradual deceleration with a smooth function, then presumably we would not find any oscillatory features in the spectrum. 

%=========
Finally, we show in \fref{fig:model3_fig4} the momentum dependence of the radiation assuming collinear kinematics \textit{i.e.} $\vvec_a$  and $\vvec_b$ remain parallel but can be arbitrarily chosen as detailed in \sref{subsub:choice1}.
Here we have numerically integrated over $\cos\theta$ to obtain the photon's momentum spectrum $p \, \dd\Pbb_{0\to0\gamma} / \dd p$.  
Toward small $p$ the curves approach a constant, which indicates that the spectrum is log-flat for small values of the momentum.  
This behavior confirms the analytical analysis that we presented in \sref{sub:model3_emission}.  
For larger values of $p$ the curves begin to decrease toward zero, which indicates that the finite wall thickness introduces a UV cutoff on the log-flat momentum spectrum.  
From the right panel it may be seen that the cutoff is parametrically given by $p_\UV \sim \gamma_w^2 / t_w$ where $\gamma_w \sim \gamma_a \sim \gamma_b$ is the Lorentz factor for the particles (assumed similar) and $t_w \sim 2 L_w / (v_a + v_b)$ is the characteristic duration of the deceleration.  
The same $\gamma^2 / t$ scaling is familiar from studies of bremstrahlung radiation in classical electromagnetism; see Ch.~15 of \rref{Jackson:1998nia}.  

%==================================
% Summary, discussion, and conclusion
%==================================
\section{Summary, discussion, and conclusion}
\label{sec:conclusion}

%=========
%Summary of what we did
To begin, we summarize the key elements of the work reported in this article.  
\begin{itemize}
    \item  We reviewed the semiclassical current radiation (SCR) formalism for calculating quantum electromagnetic radiation (photons) from a classical electromagnetic current, specifically a point-like charged particle.  
    \item  We present derivations for the vacuum survival probability $\Pbb_{0\to0}$ and the differential single emission probability $\dd\Pbb_{0\to0\gamma}$.  These expressions are valid to all orders in the electromagnetic coupling $e^2$, and they allow the probabilities to be calculated in terms of the Fourier transform of the electromagnetic current density $\tilde{j}^\mu(k)$.  These relations, which appear in eqs.~\eqref{eq:P00_from_W002},~\eqref{eq:W002_from_jj},~and~\eqref{eq:dP00g_from_jj}, are our master formulae.  We apply them to three `models' for the electromagnetic current.  
    \item  First in \sref{sec:model1}, we consider a charged particle traveling at constant velocity.  We calculate the vacuum survival probability and the differential single photon emission probability finding $\Pbb_{0\to0} = 1$ and $\dd\Pbb_{0\to0\gamma} = 0$.  As expected, there is no radiation without acceleration.  
    \item  Second in \sref{sec:model2}, we consider a charged particle that abruptly decelerates from velocity $\vvec_a$ to velocity $\vvec_b$ with $0 \leq |\vvec_b| < |\vvec_a| < 1$.  We calculate a probability distribution over the photon momentum $\pvec$ by taking $\vvec_a = \pvec_a / E_a$ and $\vvec_b = \pvec_b / E_b$ as parameters; these results appear in \erefs{eq:model2_dP00g_3vec_v}{eq:model2_dP00g_3vec_p}.  We find that the probability distribution is flat in the logarithm of the photon momentum, see \eref{eq:model2_pdPdp}.  We compare our results with the analogous calculation of bremstrahlung radiation in classical electromagnetism, finding that they lead to equivalent spectra. 
    \item  Also in \sref{sec:model2}, we consider three choices for the momentum $\pvec_b$ of the recoiling particle.  We discuss how the choice of $\pvec_b$ strongly impacts how the thermal pressure $P_\mathrm{therm}$ scales with the Lorentz factor of the bubble wall $\gamma_w$.  See the discussion of \tref{tab:summary} below for more details. 
    \item  Third in \sref{sec:model3}, we consider a gradual deceleration from velocity $\vvec_a = v_a \, \vhat$ to velocity $\vvec_b = v_b \, \vhat$ over an interval of proper time $\tau_w$, corresponding to a spatial displacement $L_w$, which we call the wall thickness.  We decompose the differential single photon emission probability into two terms.  We calculate the first term analytically, see \eref{eq:model3_pdPdp}, and we calculate both terms together numerically.  Both analyses reveal that the differential probability of photon emission is log-flat up to a momentum $p = O(\gamma_w^2/L_w)$, and for larger momenta it decreases toward zero.  
\end{itemize}

%=========
\begin{table}[t]
\begin{center}
\begin{tabular}{c|c|c|c|c|c}
%-----
formalism 
& channel 
& how to choose $\pvec_b$ 
& $\langle \Delta p_z \rangle$ 
& UV cutoff $p_\UV$ 
& $P_\mathrm{therm}$ \\ \hline \hline
%-----
C 
& $a \to b$ 
& $\circ$ 
& $\frac{m_b^2 - m_a^2}{2 E_a}$ 
& $\circ$ 
& $\propto \gamma_w^0$ \\ \hline 
%-----
QPS 
& $a \to bc$ 
& $\circ$ 
& $\frac{q^2 e^2}{4\pi^2} \Bigl( \log \frac{m_b}{m_a} - \frac{m_b^2 - m_a^2}{m_b^2 + m_a^2} \Bigr) \frac{m_b^2 + m_a^2}{E_a} \log \frac{p_\UV}{p_\IR}$ 
& $\circ$ 
& $\propto \gamma_w^0$ \\ \hline 
%-----
\multirow{3}{0.7cm}{\text{SCR}} 
& $a \to bc$ 
& 1-to-1 kinematics 
& $\frac{q^2 e^2}{4\pi^2} \Bigl( \log \frac{m_b}{m_a} - \frac{m_b^2 - m_a^2}{m_b^2 + m_a^2} \Bigr) \frac{m_b^2 + m_a^2}{E_a} \log \frac{p_\UV}{p_\IR}$ 
& $\circ$ 
& $\propto \gamma_w^0$ \\ 
%-----
%
& $a \to bc$
& 1-to-2~kinematics
& $\frac{q^2 e^2}{2 \pi^2} \, p_\UV$
& $E_a$ 
& $\propto \gamma_w^2$ 
\end{tabular}
\end{center}
\label{tab:summary}
\caption{Summary and comparison of results for the thermal pressure due to an abruptly decelerated charge.  In this article, we develop the ``semiclassical current radiation'' (SCR) formalism, and for the purpose of comparison we also present the ``classical'' (C) result of BM09, and we review the ``quantum particle splitting'' (QPS) formalism of BM17 in \aref{app:comparison}.  For the $a \to b$ channel, a particle with mass $m_a$ and energy $E_a \sim \gamma_w T$ is incident on a planar bubble wall where its mass grows to $m_b$.  For the $a \to bc$ channel, there is an additional massless radiation.  
The model with massive radiation~\cite{Bodeker:2017cim} is also reviewed in \aref{app:comparison}. 
In the SCR formalism, one must specify a prescription for choosing the recoiling particle's momentum $\pvec_b$.  If particle $c$ is soft, then it is reasonable to choose $\pvec_b$ using the 1-to-1 particle kinematics (energy and transverse momentum conservation for on-shell particles $a$ and $b$), corresponding to choice \#2 in \sref{sec:model2}.  We also show the results obtained by imposing 1-to-2 kinematics (energy and transverse momentum conservation for on-shell particles $a$, $b$, and $c$), corresponding to choice \#3 in \sref{sec:model2}.  The latter calculation results in a power law sensitivity to the UV cutoff $p_\UV$, which is expected to be the smaller than the incident particle energy $E_a \sim \gamma_w T$ (see the text for a discussion of how the nonzero bubble wall thickness $L_w$ also cuts off the momentum integration). The thermal pressure is parametrically $P_\mathrm{therm} \sim \Fcal_a \, \langle \Delta p_z \rangle$ where $\Fcal_a \sim \gamma_w T^3$ is the flux of incident thermal $a$ particles in the wall's rest frame.}
\end{table}

%=========
%Discussion:  results for abrupt deceleration
For the model with an abruptly decelerated charge, \tref{tab:summary} compares relevant results from the literature with our results derived here.  
If there is no radiation, then the calculation can be performed using a fully classical (C) formalism as in BM09~\cite{Bodeker:2009qy}.  
One finds that the longitudinal momentum transfer is $\Delta p_z \approx (m_b^2 - m_a^2) / 2E_a$ by imposing energy conservation, transverse momentum conservation, and on-shell conditions.  
In the rest frame of the bubble wall, an incident particle that is drawn from the thermal bath has energy $E_a \sim \gamma_w T$, and it follows that the thermal pressure scales as $P_\mathrm{therm} \sim \gamma_w T^3 \Delta p_z \propto \gamma_w^0$.  
To study the channel $a \to bc$ in which there is additional soft radiation, the authors of BM17~\cite{Bodeker:2017cim} employed what we call the ``quantum particle splitting'' (QPS) formalism to derive the matrix element and emission probability; these calculations are reviewed in \aref{app:comparison}.  
For the model in which a charged radiator increases mass and thereby emits a soft and massless vector (\textit{e.g.}, gluon or photon), the average longitudinal momentum transfer is found to scale as $E_a^{-1} \propto \gamma_w^{-1}$, which implies a thermal pressure $P_\mathrm{therm} \propto \gamma_w^0$ (for $m_c = 0$).  
At large $\gamma_w$ this contribution is subleading if the model also admits radiators that emit a massive vector boson (\textit{e.g.}, $W^\pm$ or $Z^0$), which give $P_\mathrm{therm} \propto \gamma_w^1$ (for $m_c \neq 0$); see \aref{app:comparison}.  

%=========
\Tref{tab:summary} also summarizes our results for the model with an abruptly decelerating charge, which we have derived using a formalism that we call ``semiclassical current radiation'' (SCR) following HKLTW20~\cite{Hoche:2020ysm}. 
Unlike the QPS formalism, which yields a joint probability distribution over both $\pvec_b$ and $\pvec$ with $\pvec_a$ as a parameter, the SCR formalism furnishes a probability distribution over $\pvec$ alone where both $\pvec_a$ and $\pvec_b$ are parameters.  
We explore several ways of choosing $\pvec_b$, which lead to different results for the average longitudinal momentum transfer $\langle \Delta p_z \rangle$ and the thermal pressure $P_\mathrm{therm}$.  
If $\pvec_b$ is chosen according to the 1-to-1 kinematics (energy and transverse momentum conservation for on-shell particles $a$ and $b$), then $\langle \Delta p_z \rangle$ is found to take exactly the same form as the expression obtained from the QPS formalism (at leading order in large $E_a$), and it follows that $P_\mathrm{therm} \propto \gamma_w^0$ (for $m_c = 0$).  
This agreement should not be surprising, since the 1-to-1 kinematics are applicable when particle $c$ is soft, and the soft splitting functions were used when deriving the expression for $\langle \Delta p_z \rangle$ in the QPS formalism.  
To reiterate, the QPS and SCR formalisms yield the same result if the phase space of the final state radiation is in the soft regime.

%=========
On the other hand, to explore the part of phase space where particle $c$ is not soft, one may instead choose to adopt the 1-to-2 kinematics (energy and transverse momentum conservation for on-shell particles $a$, $b$, and $c$), although this choice intermingles the random variable $\pvec$ with the parameter $\pvec_b$.  
With this choice $\langle \Delta p_z \rangle$ develops a linear sensitivity to the UV cutoff of the photon momentum integral $p_\UV$.  
If one takes $p_\UV \sim E_a$, corresponding to the incident particle energy, then $E_a \sim \gamma_w T$ implies $P_\mathrm{therm} \propto \gamma_w^2$, which coincides with the scaling reported in HKLTW20.  
Note that this parametric agreement is found even without the resummation of multiple soft emissions, which was performed in HKLTW20.  
Since the momentum integral appearing in the thermal pressure is UV-dominated, rather than IR-dominated, we do not expect that extending the calculation to include multiple soft emissions would significantly impact the result.
In summary, one should take care when employing the SCR formalism to study thermal pressure away from the limit where particle $c$ is soft. 

%=========
%Discussion:  results for gradual deceleration
Finally, let us discuss the conclusions that can be drawn from considering a model of gradual deceleration.  
The calculations in \sref{sec:model3} demonstrate that a finite duration of deceleration leads to a suppression of the high-$p$ emission spectrum.  
In effect, the wall thickness $L_w$ introduces an additional scale $\gamma_w^2/L_w$ that will cut off the photon momentum integral.  
This behavior is familiar from studies of electron-nucleon scattering, since the size of the nucleus enters the form factor to cut off the scattering cross section at large momentum transfer \cite{Thomson:2013zua}.  
In addition, the same $\gamma^2$ scaling arises in studies of classical bremstrahlung radiation; see Ch.~15 of \rref{Jackson:1998nia}.  

%=========
The impact of the bubble wall thickness on the thermal pressure depends on whether $\pvec_b$ is chosen according to the 1-to-1 or the 1-to-2 kinematics.  
For the $a \to bc$ channel in either QPS formalism or SCR formalism with $\pvec_b$ chosen according to the 1-to-1 kinematics, the average longitudinal momentum transfer $\langle \Delta p_z \rangle$ is only logarithmically sensitive to the UV cutoff $p_\UV$ (assuming that $p_\UV$ is large compared to the masses).  
Therefore, the gradual deceleration (\textit{i.e.}, effect of finite wall thickness) will not qualitatively change the calculation (provided that $L_w^{-1}$ remains large compared to the masses); the thermal pressure will scale as $P_\mathrm{therm} \propto \gamma_w^0$ regardless.  
If $L_w^{-1}$ becomes small compared to the masses, earlier studies have found that the thermal pressure is suppressed; see for example \rref{Gouttenoire:2021kjv}.  

%=========
On the other hand, considering the $a \to bc$ channel in SCR formalism with $\pvec_b$ chosen according to the 1-to-2 kinematics, we find that the average longitudinal momentum transfer is linearly sensitive to the UV cutoff:  $\langle \Delta p_z \rangle = q^2 e^2 p_\UV / 2\pi^2$.  
If the bubble wall thickness can be neglected, then a reasonable choice for the UV cutoff is the incident particle energy $p_\UV \sim E_a \sim \gamma_w T$.  
If the bubble wall thickness is nonzero, then it introduces an additional UV cutoff $p_\UV \propto L_w^{-1}$, which may also depend on $\gamma_w$.  
For the following estimates we take $p_\UV \sim \gamma_w^2 / L_w$, motivated by our numerical study in \sref{sec:model3} where we considered the SCR formalism with collinear kinematics; we have not calculated the spectrum using both 1-to-2 kinematics and finite wall thickness. 
For models in which $\gamma_w T < \gamma_w^2/L_w$ the incident particle's energy sets the UV cutoff, and for models in which $\gamma_w^2/L_w < \gamma_w T$ the wall thickness sets the UV cutoff.  
As for the electroweak phase transition, in models where the transition is predicted to be first order, it is typically the case that both the phase transition temperature and the wall thickness are controlled by the Higgs boson mass: $T \sim m_h$ and $L_w \sim m_h^{-1}$.  
For such models, $\gamma_w^2/L_w < \gamma_w T$ and the wall thickness provides the UV cutoff to the log-flat momentum distribution.  
Then the average longitudinal momentum transfer is expected to be parametrically $\langle \Delta p_z \rangle \sim \gamma_w^2 e^2 / L_w$, and the thermal pressure exerted by the plasma on the wall is expected to be parametrically $P_\mathrm{therm} \sim \gamma_w T^3 \, \langle \Delta p_z \rangle \sim e^2 \gamma_w^3 T^3 / L_w$.  
The same scaling with $e$ and $L_w$ was obtained in BM17 using a different method; see their eq.~(5.7), although the scaling with mass is different.  
By contrast, for models in which the ordering of the UV scales is reversed such that $\gamma_w T < \gamma_w^2/L_w$, the thermal pressure is instead expected to have the parametric scaling $P_\mathrm{therm} \sim e^2 \gamma_w^2 T^4$.  
This would rise more quickly with increasing $\gamma_w$, although it still cannot exceed $e^2 \gamma_w^3 T^3 / L_w$ since the derivation assumed $\gamma_w T < \gamma_w^2/L_w$.  

%======================================
\section*{Acknowledgments}
We are grateful to Stefan H{\"o}che for collaboration during the early stages of this project and to Yuber Perez-Gonzalez and Bibhushan Shakya for constructive feedback.
J.T. is supported by the STFC under Grant No.~ST/T001011/1.
A.J.L. is partly supported by the NASA ATP award 80NSSC22K0825.   
A.J.L. is grateful to Durham University for their hospitality while a portion of this work was being performed, and to the DIVA Program whose financial award made this collaboration possible.  

%======================================
%======================================
\begin{appendix}

%==================================
% Cancellation of IR divergences
%==================================
\section{Cancellation of IR divergences}
\label{app:cancellation}

%=========
In this appendix we illustrate how IR divergences cancel when calculating the probability for no detectable emission.  
Summarizing the results of sections~\ref{sec:vacuum}~and~\ref{sec:emission}, the vacuum survival probability is 
\bes{
    \Pbb_{0\to0} = \mathrm{exp}\Bigl( \mathrm{Re} \, W_{0\to0}^{(2)} \Bigr) 
    \qquad \text{with} \qquad 
	W_{0\to0}^{(2)} = \lim_{\epsF \to 0^+} \int \! \! \frac{\dd^4 k}{(2\pi)^4} \, \frac{\ii}{k \cdot k + \ii \epsF} \, \tilde{j}(k)^\ast \cdot \tilde{j}(k) 
    \com
}
and the single emission probability is 
\ba{
    \Pbb_{0\to0\gamma}
	= - \Pbb_{0\to0} \int \! \! \frac{\dd^3 \pvec}{(2\pi)^3} \, \frac{1}{2E_p} \, \tilde{j}(p)^\ast \cdot \tilde{j}(p) 
    \com
}
where $p^\mu = (E_p, \pvec)$ and $E_p = |\pvec|$.  
If $\tilde{j}(k)$ doesn't have any poles and if it decreases sufficiently quickly toward large $|k^0|$ then the integral over $k^0 = \omega$ can be evaluated using the Residue Theorem.  
Closing the integration contour in the lower half plane leads to 
\ba{
    W_{0\to0}^{(2)} 
    & = \int \! \! \frac{\dd^3 \kvec}{(2\pi)^3} \, \frac{1}{2\omega_k} \, \tilde{j}(k)^\ast \cdot \tilde{j}(k) 
    \com
}
where $k^\mu = (\omega_k, \kvec)$ and $\omega_k = |\kvec|$.  
When written in this way, it is clear that the vacuum persistence amplitude $W_{0\to0}^{(2)}$ is real.  
Notice that $\Pbb_{0\to0\gamma}$ also contains a factor of $W_{0\to0}^{(2)}$, which only differs by the relabeling $\kvec \to \pvec$ and $\omega_k \to E_p$.  
Consequently the single emission probability can also be written as 
\ba{
    \Pbb_{0\to0\gamma}
	= - \Pbb_{0\to0} \ \mathrm{Re} \, W_{0\to0}^{(2)} 
    \com
}
where we have already noted that $\mathrm{Re} \, W_{0\to0}^{(2)}$ must be negative to ensure a resonable probability.  

%=========
Suppose we are interested in the probability that no detectable emission occurs: 
\ba{
    \Pbb_\text{no detectable emission} 
	& = \Pbb_{0\to0} + \Pbb_{0\to0\gamma}^\text{soft/col} + \Pbb_{0\to0\gamma\gamma}^\text{soft/col} + \cdots 
    \per
}
This quantity is the sum of the probability that no emission occurs $\Pbb_{0\to0}$, the probability that a single soft or collinear photon is emitted $\Pbb_{0\to0\gamma}^\text{soft/col}$, and the probability for emitting several such photons.  
Photons that are too soft or too collinear with the radiating particle cannot be detected, and so they contribute to the probability for no detectable emission.  
When calculating $\Pbb_{0\to0}$ we encountered a logarithmic IR divergence, and we expect that this singularity is cancelled by $\Pbb_{0\to0\gamma}^\text{soft/col}$ such that the measurable quantity $\Pbb_\text{no detectable emission}$ is singularity-free.  
To understand how this cancellation might occur, let us calculate the sum of $\Pbb_{0\to0}$ and $\Pbb_{0\to0\gamma}$ without singling out the soft/collinear parts: 
\bes{
    \Pbb_{0\to0} + \Pbb_{0\to0\gamma} 
	& = \Pbb_{0\to0} + \bigl( - \Pbb_{0\to0} \ \mathrm{Re} \, W_{0\to0}^{(2)} \bigr) \\ 
	& = \Bigl( 1 + \mathrm{Re}\, W_{0\to0}^{(2)} + O(e^4) \Bigr) \Bigl( 1 - \mathrm{Re}\, W_{0\to0}^{(2)} \Bigr) \\ 
	& = 1 + O(e^4)
	\per
}
Here we see that there is a cancellation between the virtual corrections and the real emission at $O(e^2)$.  
This suggests that the measurable probability for no detectable emission is free of IR divergences.  

%==================================
% Comparison with fully quantum formalism
%==================================
\section{Comparison with quantum particle splitting formalism}
\label{app:comparison}

%=========
In this work, we have used the semiclassical current radiation formalism to calculate the spectrum of photon emission that arises when a charged particle decelerates at a planar bubble wall.  
In this appendix, we compare with the quantum particle splitting formalism developed in BM17~\cite{Bodeker:2017cim}.

%=========
Suppose that a particle of species $a$ is incident on the wall, and particles of species $b$ and $c$ are recoiling.  
Let the masses in the symmetric and Higgs phases be denoted by $m_{f,\mathrm{s}}$ and $m_{f,\mathrm{h}}$ for $f=a,b,c$.
If particle $a$ is in the one-particle state labeled by momentum $\pvec_{a,\mathrm{s}}$, then the differential probability for particles $b$ and $c$ to recoil into the two-particle state labeled by momenta $\pvec_{b,\mathrm{s}}$ and $\pvec_{c,\mathrm{s}}$ may be calculated using eqs.~(3.3)~and~(3.15) of HKLTW20 (see also eq.~(3.10) of BM17)
\bes{
    \dd \Pbb_{a\to bc} = \frac{1}{2E_a} \frac{\dd^3 \pvec_{b,\mathrm{s}}}{(2\pi)^3} \frac{1}{2E_b} \frac{\dd^3 \pvec_{c,\mathrm{s}}}{(2\pi)^3} \frac{1}{2E_c} \, 
    (2\pi)^3 \, \frac{p_{a,z,\mathrm{s}}}{E_a} \, 
    \delta(\pvec_{a,\perp} - \pvec_{b,\perp} - \pvec_{c,\perp}) \, 
    \delta(E_a - E_b - E_c) \, 
    |\Mcal_{a\to bc}|^2 
    \per
}
The kinematic variables are $\pvec_{f,\mathrm{s}} = (\pvec_{f,\perp}, p_{f,z,\mathrm{s}})$, $p_{f,z,\mathrm{s}} = (E_f^2 - |\pvec_{f,\perp}|^2 - m_{f,\mathrm{s}}^2)^{1/2}$, and $p_{f,z,\mathrm{h}} = (E_f^2 - |\pvec_{f,\perp}|^2 - m_{f,\mathrm{h}}^2)^{1/2}$ for $f=a,b,c$.  
Evaluating the integral over $\pvec_{b,\mathrm{s}}$ yields the probability distribution over $\pvec_{c,\mathrm{s}}$: 
\bes{
    \dd \Pbb_{a\to bc} = \frac{\dd^3 \pvec_{c,\mathrm{s}}}{(2\pi)^3} \, \frac{1}{8 E_a E_b E_c} \, |\Mcal_{a\to bc}|^2 \Bigr|_{\pvec_{b,\perp} = \pvec_{a,\perp} - \pvec_{c,\perp}, E_b = E_a - E_c}
    \per
}
We work in the frame where $\pvec_{a,\perp} = 0$.  
We define $\kvec_\perp \equiv \pvec_{c,\perp}$, and transverse momentum conservation implies $\pvec_{b,\perp} = - \kvec_\perp$. 
We further define $x = E_c / E_a$, and energy conservation implies $E_b = (1-x) E_a$.  
After a change of variables, the probability distribution is expressed as 
\bes{\label{eq:dPabc_dkP_dx}
    \dd \Pbb_{a\to bc} = \dd k_\perp \, \dd x \, \frac{k_\perp}{32 \pi^2 (1-x) E_a \sqrt{x^2 E_a^2 - k_\perp^2 - m_{c,\mathrm{s}}^2}} \, |\Mcal_{a\to bc}|^2 \Bigr|_{\pvec_{b,\perp} = \pvec_{a,\perp} - \pvec_{c,\perp}, E_b = E_a - E_c}
    \com
}
where $k_\perp = |\kvec_\perp|$ is the magnitude of the transverse momentum. 

%=========
In the QPS formalism, the matrix element $\Mcal_{a\to bc}$ may be calculated using the WKB approximation.  
If the thickness of the wall is neglected, then an expression for the matrix element appears in eq.~(3.18) of HKLTW20 (see also eq.~(4.8) of BM17)
\bes{
    \Mcal_{a\to bc}^{(0)} = 2 \ii E_a \biggl( \frac{V_\mathrm{h}}{A_\mathrm{h}} - \frac{V_\mathrm{s}}{A_\mathrm{s}} \biggr) 
    \com
}
where $V_\mathrm{h}$ and $V_\mathrm{s}$ are vertex functions and where 
\bes{
    A_\mathrm{s} & = - 2 E_a \, \bigl( p_{a,z,\mathrm{s}} - p_{b,z,\mathrm{s}} - p_{c,z,\mathrm{s}} \bigr) \\ 
    A_\mathrm{h} & = - 2 E_a \, \bigl( p_{a,z,\mathrm{h}} - p_{b,z,\mathrm{h}} - p_{c,z,\mathrm{h}} \bigr) 
    \per
}
Particles $a$ and $b$ are assumed to be charged radiators (scalar, spinor, or vector), and particle $c$ is assumed to be a transversely-polarized spin-$1$ boson. 
In the part of phase space where particle $c$ is soft ($0 < x \ll 1$), the squared splitting functions can be approximated as (see table~1 of BM17)
\bes{
    |V_\mathrm{s}|^2 = |V_\mathrm{h}|^2 = 4 g^2 C_2[R] \, \frac{k_\perp^2}{x^2} 
    \com
}
where $g$ is a coupling and $C_2[R]$ is the quadratic Casimir for the incident particles ($C_2[R] = 1$ for electromagnetism).  

%=========
The observable of interest is the average longitudinal momentum transfer: 
\bes{
    \langle \Delta p_z \rangle = \int \! \dd \Pbb_{a\to bc} \, \Delta p_z 
    \qquad \text{where} \qquad 
    \Delta p_z = p_{a,z,\mathrm{s}} - p_{b,z,\mathrm{h}} - p_{c,z,\mathrm{h}} 
    \per
}
Expanding for large $E_a$ while $x$, $k_\perp$, and the masses are held fixed leads to 
\bes{\label{eq:Delta_pz_appendix}
    \Delta p_z \approx \biggl( \frac{k_\perp^2 + m_{c,\mathrm{h}}^2}{2 x} + \frac{k_\perp^2 + m_{b,\mathrm{h}}^2}{2(1-x)} - \frac{m_{a,\mathrm{s}}^2}{2} \biggr) \frac{1}{E_a} + O(E_a^{-3}) 
    \per
}
This approximation corresponds to the assumption that each particle's longitudinal momentum is large compared to its transverse momentum and its mass. 

%=========
Now we consider two cases.  
In order to compare with the main results of BM17, we first suppose that the radiator particles remain massless, and the mass of the emitted vector changes across the wall.  
This corresponds to setting $m_{a,\mathrm{s}} = m_{a,\mathrm{h}} = m_{b,\mathrm{s}} = m_{b,\mathrm{h}} = 0$.  
The squared matrix element evaluates to 
\bes{
    |\Mcal_{a\to bc}^{(0)}|^2 \approx 16 g^2 C_2[R] \frac{k_\perp^2 \, \bigl( m_{c,\mathrm{h}}^2 - m_{c,\mathrm{s}}^2 \bigr)^2 \, (1-x)^4}{\bigl( k_\perp^2 + (1-x)^2 \, m_{c,\mathrm{h}}^2 \bigr)^2 \bigl( k_\perp^2 + (1-x)^2 \, m_{c,\mathrm{s}}^2 \bigr)^2} \, E_a^2 
    +O(E_a^{0})
    \per
}
Here we have expanded for large $E_a$ while holding fixed $x$, $k_\perp$, and the masses.  
Setting $x=0$ yields an expression that matches Eq.~(3.21) of HKLTW20 (or Eq.~(4.13) of BM17 if $m_{c,\mathrm{s}} = 0$).  
The corresponding differential probability can be evaluated from \eref{eq:dPabc_dkP_dx}.  
For a fixed $x$, the integrand is log-flat across the region $(1-x) m_{c,\mathrm{s}} < k_\perp < (1-x) m_{c,\mathrm{h}}$, which implies that $k_\perp$ will be on the order of the emitted particle's mass.  
Evaluating the integral over $k_\perp$ leads to the differential probability for emitting a $c$ particle with energy fraction between $\ln x$ and $\ln x + \dd \ln x$: 
\bes{\label{eq:dP_case1}
    x \frac{\dd \Pbb_{a\to bc}}{\dd x} & \approx 
    \frac{g^2 C_2[R]}{2 \pi^2} 
    \biggl( \frac{m_{c,\mathrm{h}}^2 + m_{c,\mathrm{s}}^2}{m_{c,\mathrm{h}}^2 - m_{c,\mathrm{s}}^2} \log \frac{m_{c,\mathrm{h}}}{m_{c,\mathrm{s}}} - 1 \biggr) \, 
    \bigl( 1-x \bigr) 
    + O(E_a^{-2})
    \per
}
If the masses were hierarchical, $m_{c,\mathrm{s}}^2 \ll m_{c,\mathrm{h}}^2$, then the logarithm would be large, as a consequence of the would-be collinear divergence IR divergence at $\kvec_\perp = 0$.  
The spectrum is log-flat in the energy of the recoiling particle ($E_c = x E_a$) provided that it is small compared to the incident particle's energy.  
As $x$ approaches $1$ from below, the formula breaks down, and the emission is suppressed.  
To calculate the average longitudinal momentum transfer, since $\Delta p_z \propto m_{c,\mathrm{h}}^2/x E_a$ in \eref{eq:Delta_pz_appendix}, there is a power law IR divergence.  
Integrating $k_\perp$ from $0$ to $\infty$ and $x$ from $x = x_\IR$ to $x = x_\UV$ gives 
\bes{\label{eq:Delta_pz_massive_emission}
    \langle \Delta p_z \rangle \approx 
    \frac{g^2 C_2[R]}{4 \pi^2} 
    \biggl( \log \frac{m_{c,\mathrm{h}}}{m_{c,\mathrm{s}}} - \frac{m_{c,\mathrm{h}}^2 - m_{c,\mathrm{s}}^2}{2 m_{c,\mathrm{h}}^2} \biggr) \frac{m_{c,\mathrm{h}}^2}{E_a \, x_\IR} 
    + O(E_a^{-3}) 
    \com
}
where we have dropped terms that are subleading for $0 < x_\IR \ll x_\UV < 1$.  
Note that power-law singularity toward small $x$ has led to a factor of $E_a x_\IR = E_{c,\IR}$ in the denominator which is on the order of $m_{c,\mathrm{s}}$ or $m_{c,\mathrm{h}}$.  
So if the masses differ by an $O(1)$ factor then the average longitudinal momentum transfer is parametrically $\langle \Delta p_z \rangle \sim g^2 m_{c}$ and the thermal pressure is parametrically $P_\mathrm{therm} \sim g^2 m_c \gamma_w T^3 \propto \gamma_w^1$, which is one of the main results of BM17.  

%=========
Now we consider the second case.  
In order to compare with our results in the main body of the paper, we suppose that the radiator particle develops a mass as it crosses the wall but the radiation remains massless.  
This corresponds to setting $m_{b,\mathrm{s}} = m_{a,\mathrm{s}}$, $m_{a,\mathrm{h}} = m_{b,\mathrm{h}}$, and $m_{c,\mathrm{s}} = m_{c,\mathrm{h}} = 0$.  
The squared matrix element evaluates to 
\bes{
    |\Mcal_{a\to bc}^{(0)}|^2 \approx 16 g^2 C_2[R] \frac{k_\perp^2 \, \bigl( m_{b,\mathrm{h}}^2 - m_{a,\mathrm{s}}^2 \bigr)^2 \, (1-x)^2 x^4}{\bigl( k_\perp^2 + x^2 \, m_{b,\mathrm{h}}^2 \bigr)^2 \bigl( k_\perp^2 + x^2 \, m_{a,\mathrm{s}}^2 \bigr)^2} \, E_a^2 
    +O(E_a^{0})
    \per
}
Note the additional factor of $x^4$, which will tend to suppress the matrix element in the region of phase space where the radiated particle is soft ($x = E_c / E_a \ll 1$).  
Now the integral over $k_\perp$ is log-flat across the region $x m_{a,\mathrm{s}} < k_\perp < x m_{b,\mathrm{h}}$, and if $x \ll 1$ the typical transverse momentum is small compared to the masses.  
Evaluating the $k_\perp$ integral yields the energy spectrum of the resultant radiation 
\bes{\label{eq:dP_case2}
    x \frac{\dd \Pbb_{a\to bc}}{\dd x} & \approx 
    \frac{g^2 C_2[R]}{2 \pi^2} 
    \biggl( \frac{m_{b,\mathrm{h}}^2 + m_{a,\mathrm{s}}^2}{m_{b,\mathrm{h}}^2 - m_{a,\mathrm{s}}^2} \log \frac{m_{b,\mathrm{h}}}{m_{a,\mathrm{s}}} - 1 \biggr) \, 
    \bigl( 1-x \bigr) 
    + O(E_a^{-2})
    \per
}
Note that \eref{eq:dP_case1} displays the same log-flat energy spectrum that we already encountered in \eref{eq:dP_case2}, despite the additional powers of $x^4$ that now appear in the squared matrix element.  
This is because the domain of the $k_\perp$ integration is restricted to $k_\perp \sim x m_a$ where the integrand is effectively enhanced by a factor of $1/x^4$.  
Perhaps more importantly, it is worth emphasizing that \eref{eq:dP_case2} is equivalent to the energy spectrum that we derived using the SCR formalism when studying the abrupt deceleration model in \sref{sec:model2} and selecting the incident and recoiling particle velocities to match the 1-to-1 kinematics; compare with \eref{eq:model2_pdPdp_choice2}.  
This agreement goes toward validating the SCR formalism as a reliable description of the system. 
Finally we calculate the average longitudinal momentum transfer to find 
\bes{\label{eq:Delta_pz_massless_emission}
    \langle \Delta p_z \rangle \approx 
    \frac{g^2 C_2[R]}{4 \pi^2} 
    \biggl( \log \frac{m_{b,\mathrm{h}}}{m_{a,\mathrm{s}}} - \frac{m_{b,\mathrm{h}}^2 - m_{a,\mathrm{s}}^2}{m_{b,\mathrm{h}}^2 + m_{a,\mathrm{s}}^2} \biggr) \frac{m_{b,\mathrm{h}}^2 + m_{a,\mathrm{s}}^2}{E_a} \log \frac{x_\UV}{x_\IR} 
    + O(E_a^{-3}) 
    \com
}
where we have dropped terms that are subleading for $0 < x_\IR \ll x_\UV < 1$.  
Now since $m_{c,\mathrm{h}} = 0$ and since $k_\perp$ is predominantly supported on the interval from $x m_{a,\mathrm{s}}$ to $x m_{b,\mathrm{h}}$, the expression for $\Delta p_z$ is not singular at $x = 0$.  
So the power law dependence on $x_\IR$ is softened to a logarithmic dependence.  
We note that the average momentum transfer shown above in 
\eqref{eq:Delta_pz_massless_emission}, calculated  using the QPS formalism, is the same as the momentum transfer computed using the SCR formalism \eqref{eq:model2_choice2_Dpz} if we assume 1-to-1 kinematics, \textit{i.e.} that the decelerating particle's transverse momentum does not change significantly after passing the bubble wall.

%=========
It is illuminating to compare the expressions for $\langle \Delta p_z \rangle$ in \erefs{eq:Delta_pz_massive_emission}{eq:Delta_pz_massless_emission}.  
If the emission is massive, then the average longitudinal momentum transfer goes as $1 / E_a x_\IR \sim 1 / m_c$ as in \eref{eq:Delta_pz_massive_emission}, whereas if the emission is massless then it goes as $1 / E_a$ as in \eref{eq:Delta_pz_massless_emission}.  
In order to track the $\gamma_w$ dependence, note that $E_a \sim \gamma_w T$. 
For the model with massive vector boson emission $\langle \Delta p_z \rangle \propto E_a^0 \propto \gamma_w^0$.  
However, for the model with massless vector boson emission and a massive radiator $\langle \Delta p_z \rangle \propto E_a^{-1} \propto \gamma_w^{-1}$. 
As noted by BM17, if the model (\textit{e.g.}, electroweak standard model) allows for emission via both channels, then the latter channel is subleading for large $\gamma_w$.  

\end{appendix}

\bibliographystyle{JHEP}
\bibliography{refs}

\end{document}